\documentclass[numberedappendix]{emulateapj}
\usepackage[hyperfootnotes=false,naturalnames=true,letterpaper,pdfstartview=FitH,pdfpagemode=none]{hyperref}
\hypersetup{pdfauthor={Dan Coe}, pdftitle={LensPerfect A1689}}
\let\chapter=\section  
\usepackage{placeins}  
\usepackage{amsmath}
\usepackage{rotating}

\newcommand{\gammainc}{\hat\gamma}

\newcommand{\gp}{\gamma^\prime}

\slugcomment{Draft}

\begin{document}


\title{Dark Matter Halo Mass Profiles}
\author{Dan~Coe\altaffilmark{1}}
\altaffiltext{1}{
  Jet Propulsion Laboratory, California Institute of Technology, 
  4800 Oak Grove Dr, MS 169-327, Pasadena, CA 91109
}

\begin{abstract}
I provide notes on the NFW, Einasto, S\'ersic, and other mass profiles
which provide good fits to simulated dark matter halos (\S\ref{massprofiles}).
I summarize various published $c(M)$ relations:
halo concentration as a function of mass (\S\ref{S:cM}).
The definition of the virial radius is discussed
and relations are given to convert $c_{vir}$, $M_{vir}$, and $r_{vir}$
between various defined values of the halo overdensity (\S\ref{virial}).
\end{abstract}




\vspace{0.2in}
\section{Mass-Concentration Relations}
\label{S:cM}

The mass profiles of galaxy clusters appear to be more centrally concentrated
than realized in simulations
\citep{Broadhurst08,Oguri09,Sereno10}.
If true, this may be evidence for Early Dark Energy \citep[see e.g.,][]{GrossiSpringel09}.
Or perhaps there is a less exciting explanation
\citep[e.g.,][]{BarkanaLoeb09,LapiCavaliere09,Meneghetti10}.
For more details, see my discussion in \cite{Coe10}.
More conclusive results are expected from 
the CLASH HST MCT project\footnote{Cluster 
Lensing And Supernova survey with Hubble,
\href{http://www.stsci.edu/~postman/CLASH/}{http://www.stsci.edu/$\sim$postman/CLASH/}
}
and perhaps LoCuSS \citep{Okabe09,Richard10}.

More massive halos generally have lower concentrations than less massive halos.
This is seen in both simulations and observations,
though less clearly so far in the latter (see below).
More massive halos form later, resulting in lower concentrations
reflecting the lower background density at the time of formation \cite{NFW96}.

For a given radial mass profile (see \S\ref{massprofiles}), 
the concentration is defined as:
\\
\begin{equation}
  c _{vir} = r_{vir} / r_{-2},
\end{equation}
\\
a mishmash ratio of the virial radius $r_{vir}$ and the radius $r_{-2}$ at which $\rho \propto r^{-2}$.
For an NFW profile, $r_{-2} = r_s$.
The definition of the virial radius $r_{vir}$ is discussed at length in \S\ref{virial},
but it is typically approximated as the region within which there is 
an average overdensity of a certain value ($\Delta_c \sim100$ or 200)
above $\rho_{crit}$.  
For clarity, one may quote the exact value of $\Delta_c$ used: $c_{200}$, for example.

In principle concentrations could be derived 
using any radial fitting profile (\S\ref{massprofiles}).
However the choice does matter 
as the profiles behave differently between $r_{-2}$ and $r_{vir}$.
Concentrations derived from NFW and Einasto fits to the same halos \citep{Duffy08}
are compared in Fig.~\ref{cMzEinrel}.
Einasto $c(M)$ relations have been derived for the Millennium simulation
relaxed \citep{Gao08} and all \cite{HayashiWhite08} halos.
Below we focus on $c(M)$ relations derived from NFW fits.

\subsection{Current $c(M)$ measurements from NFW profile fits}

The current best estimates for $c(M,z)$
are probably those given by \cite{Duffy08} and \cite{Maccio08}.
Their findings are similar.
Both analyze simulations which use the WMAP5 cosmology,
resulting in $\sim 20\%$ lower concentrations than WMAP1 (Table \ref{tab:cosmo})
as used in the Millennium simulation \citep{Neto07}, for example.
\cite{Duffy08} find that present-day ($z=0$) halos 
follow the following mass-concentration relation:
\\
\begin{equation}
  \label{Duffy08}
  c_{200} \simeq 5.74
  \left( \frac{M_{200}}{2 \times 10^{12} h^{-1} M_\odot} \right) ^ {-0.097}.
\end{equation}
\\
They provide a separate relation for relaxed clusters
which are more symmetric and thus better fit by radial profiles such as NFW.
These have $15-20\%$ higher concentrations (Fig.~\ref{crelrat}):
\\
\begin{equation}
  \label{Duffy08}
  c_{200} \simeq 6.67
  \left( \frac{M_{200}}{2 \times 10^{12} h^{-1} M_\odot} \right) ^ {-0.092}.
\end{equation}
\\
Intrinsic scatters are $\Delta \log_{10} (c_{200}) \simeq 0.15$.
These relations are plotted in Fig.~\ref{cMDuffyMaccio} 
along with corresponding relations from \cite{Maccio08}.

\cite{Duffy08} also supply fitted functions 
to halos spanning the redshift range $z = 0 - 2$.
Full:
\begin{equation}
  \label{Duffy08}
  c_{200} \simeq \frac{5.71}{(1+z)^{0.47}}
  \left( \frac{M_{200}}{2 \times 10^{12} h^{-1} M_\odot} \right) ^ {-0.097}
\end{equation}
\\
and relaxed:
\\
\begin{equation}
  \label{Duffy08}
  c_{200} \simeq \frac{6.71}{(1+z)^{0.44}}
  \left( \frac{M_{200}}{2 \times 10^{12} h^{-1} M_\odot} \right) ^ {-0.092}.
\end{equation}
\\
In their Table 1, they provide uncertainties for these fit parameters
as well as corresponding values for $c_{vir}$ and $M_{vir}$.
These $c_{200}(M_{200},z)$ relations are plotted in Figs.~\ref{cMzM} \& \ref{cMzz}.
Also plotted are the corresponding $c_{vir}(M_{vir},z)$ relations
provided by \citep{Duffy08}.
In Fig.~\ref{cMzz}, we plot the \cite{Bullock01} $c \propto (1+z)^{-1}$ scaling
for comparison.
Note that Duffy08 find weaker dependencies on redshift:
$c_{200} \propto (1+z)^{-0.45}$;
$c_{vir} \propto (1+z)^{-0.70}$.

Halo concentrations are sensitive to cosmology.
A higher $\sigma_8$ causes halos to form earlier, resulting in higher concentrations.
This was the case in the Millennium simulations 
which used the WMAP 1-year cosmology, including $\sigma_8 = 0.9$.
This yields concentrations $\sim 20\%$ higher than
found in simulations which use WMAP5's $\sigma_8 = 0.796$\footnote{This 
value is in excellent agreement with the WMAP 7-year maximum likelihood value $\sigma_8 = 0.803$ \citep{WMAP7}.} \citep{Duffy08}.
The effect of cosmology was explored in more detail by \cite{Maccio08}.
These effects are shown in Fig.~\ref{cMcosmo}.

Various derived $c(M)$ relations (for $z=0$) are plotted in Fig.~\ref{cMcompare}.
In our Tables \ref{tab:cM}, \ref{tab:cMvir}, and \ref{tab:cMz},
we provide $c_{200}(M_{200})$, $c_{vir}(M_{vir})$, and $c(M,z)$ relations, respectively,
as derived by \cite{Duffy08}, \cite{Maccio08}, \cite{Neto07},
\cite{Bullock01}, \cite{Hennawi07}, and \cite{Gentile07}.\footnote{I'll take a crack at completeness, and mention other papers with $c(M)$ relations neglected here (for no particular reason): \cite{Eke01,Wechsler02,Alam02,Zhao03b,Dolag04,Kuhlen05,Lu06,Shaw06,Dutton07,Gnedin07}.  
And I have probably missed still others!}
The latter is the original NFW $c(M)$ prescription updated to the WMAP3 cosmology.

The various simulations considered here are outlined in Table~\ref{tab:sims}.
The relevant details of their adopted cosmologies ($\Omega_m$, $\sigma_8$)
are given in Table~\ref{tab:cosmo}.
We provide the range of halo masses produced in each simulation.
The dangers of extrapolating $c(M)$ relations beyond these ranges
have been cited by \cite{Zhao03b}, for example.

\citet[their Fig.~5]{Neto07} find that 10,000 particles within the virial radius
are required to yield robust concentration measurements.
They note that using fewer particles introduces scatter
but does not appear to introduce bias in their concentration measurements.

\cite{Hennawi07} measure significantly larger concentrations
for galaxy clusters in their simulations.
Their cluster concentrations are
$\sim 50\%$ and $\sim 80\%$ larger than found by
\cite{Duffy08} and \cite{Maccio08}, respectively (Fig.~\ref{cMcompare}, {\it right}).
Their use of $\sigma_8 = 0.95$ probably only results in concentrations
inflated by $\sim 20\%$ compared to the WMAP5 $\sigma_8 = 0.796$ simulations.
The remaining disagreement may be a result of their halo density fitting procedure
which they claim is better for comparison with lensing measurements.
Specifically, they assign large uncertainties to radial bins with large subhalos.
This may bias the fitted profiles to be low at large radius 
(where large subhalos typically reside)
resulting in higher concentrations.
These results may considerably ease tensions 
between observed and simulated halo concentrations.
The differences in fitting procedure should be better studied and understood.

\subsection{Care in citing concentration expectations}

A concern often noted is that the concentration measured for A1689
(in every study to date)
is higher than that found in simulations for a halo of A1689's mass.
The concentration found in simulations
has been cited loosely as $c \sim 5$ or $c \sim 5.5$ 
using a relation given by \cite{Bullock01}:
\\
\begin{equation}
  \label{NFWcBullock01}
  c_{vir} \simeq \frac{9}{1+z}
  \left( \frac{M_{vir}}{1.3 \times 10^{13} h^{-1} M_\odot} \right) ^ {-0.13},
\end{equation}
But the expected concentration is actually lower 
($c_{200} \sim 3.0$, exacerbating the disagreement with observations) 
for four reasons:
\vspace{-0.08in}
\begin{itemize}
\item $c_{200} < c_{vir}$
\vspace{-0.08in}
\item $z > 0$ 
\vspace{-0.08in}
\item WMAP7 vs.~WMAP1
\vspace{-0.08in}
\item $M_{vir} \approx 1.4 \times 10^{15} M_\odot / h > 10^{15} M_\odot$
\end{itemize}
\vspace{-0.05in}
We also note that \cite{Bullock01} did not simulate halos as massive as A1689,
with their most massive halos weighing in at $M \sim 10^{14} M_\odot / h$.

\subsection{The observed $c(M)$ relation}

Based on a compilation of 62 published measurements 
(including 10 new measurements) of halo $c_{vir}$ and $M_{vir}$,
\cite{ComerfordNatarajan07} find the following relation (with a large scatter):
\\
\begin{equation}
  \label{NFWcComerford07}
  c_{vir} \simeq \frac{14.5 \pm 6.4}{1+z}
  \left( \frac{M_{vir}}{1.3 \times 10^{13} h^{-1} M_\odot} \right) ^ {-0.15 \pm 0.13}.
\end{equation}
\\
For clusters as massive as A1689, 
Comerford's relation converges toward that of \cite{Hennawi07},
the former being only slightly higher.

This and other observed $c(M)$ relations are shown in 
Fig.~\ref{cMcompare} and detailed in Table~\ref{tab:cMobs}.
The \cite{ComerfordNatarajan07} compilation includes 
both lensing and X-ray determinations of $c$ and $M$,
including the X-ray samples presented by \cite{Buote07} and \cite{SchmidtAllen07}.
Each of these papers presented their own $c(M)$ relation.
A recent $c(M)$ relation from weak lensing of individual halos 
was presented by \cite{Okabe09}.
And $c(M)$ derived from {\it stacked} weak lensing analyses
were presented by \cite{Johnston07} and \cite{Mandelbaum08}.
It seems apparent that one should study a wide enough range of halo masses
to obtain a confident $c(M)$ relation.

\vspace{0.2in}
\section{Overdensity within the Virial Radius}
\label{virial}

Various conventions are used to define the virial mass and radius.
We explain and show how to convert between different definitions.

\subsection{Overdensity Definitions}

The virial radius $r_{vir}$ designates the edge of the halo.
Within this radius, objects are supposed to be ``virialized'':
gravitationally bound and settled into regular orbits.
Outside this radius, objects are not in orbit although they may still be infalling.
In practice, there is no sharp dividing line the two regions.
And even if there were, it would be extremely difficult to discern 
observationally for a given massive body.
Meanwhile, the objects we study are not always virialized.
In fact, galaxy clusters are the largest bodies 
which have had time to virialize given the age of the universe.
Thus some of the clusters we observe have virialized just recently,
but many are still in the process of doing so.

Despite these complications, 
we can define a virial radius for a massive body
based on theory and simulations.
Early theoretical work \citep{Peebles80}
predicted that a sphere of material will collapse 
if its density exceeds $1.686(1+z)$ times that of the background.
After it collapses and virializes,
the sphere will obtain an average density
\begin{equation}
  \Delta_c \approx 18 \pi^2 \approx 178
\end{equation}
times the critical density $\rho_{crit}(z)$ 
at that redshift, where
\\
\begin{equation}
  \rho_{crit} = \frac{3 H^2(z)}{8 \pi G}.
\end{equation}
\\
\cite{ColeLacey96} cited this as a theoretical result
and then confirmed it in simulations.\footnote{\cite{ColeLacey96} 
spoke of this factor as the overdensity above $\rho_m = \Omega_m \rho_{crit}$ 
rather than above $\rho_{crit}$.
But as $\Omega_m = 1$ in their simulations,
the two densities were equal and thus interchangeable.}

\cite{NFW96} adopted the nice round number of $\Delta_c = 200$,
which has been used commonly ever since to allow for easy comparison
between papers.
But the $\Delta_c \approx 178$ result was obtained in an Einstein de-Sitter
cosmology of ($\Omega_m$, $\Omega_\Lambda$) = (1, 0).
In the concordance cosmology ($\Omega_m$, $\Omega_\Lambda$) = (0.3, 0.7),
we find a much lower value of $\Delta_c \approx 100$, as we describe next.

At least three different forms have been given for $\Delta_c$
as a function of cosmology.  
For a flat universe ($\Omega_m + \Omega_\Lambda = 1$),
\cite{BryanNorman98} give
\begin{equation}
  \label{BryanNorman98}
  \Delta_c \approx 18 \pi^2 - 82 \Omega_\Lambda - 39 \Omega_\Lambda^2.
\end{equation}
An approximation to this is given as \citep{Eke98}
\begin{equation}
  \Delta_c \approx 178 \Omega_m^{0.45}.
\end{equation}
And 
\citet[their Eq.~C19; see also \citealt{Henry00}, their Eq.~A17]{NakamuraSuto97}
give
\begin{equation}
  \Delta_c \approx 18 \pi^2 (1 + 0.4093 x^{2.7152}) \Omega_m
\end{equation}
with $x \equiv (1 / \Omega_{m,0} - 1)^{1/3} (1 + z)^{-1}$
and $\Omega_m(z) = 1 / (1 + x^3)$.
Given the current concordance cosmology with $\Omega_m = 0.3$,
these different expressions yield
$\Delta_c = 101.1, 103.5, 100.3$, respectively
for a halo at $z = 0$.
Or given $\Omega_m = 0.25$, $\Delta_c = 94.2, 95.4, 93.5$.
We note $\Omega_m = 0.25$ is
in better agreement with the WMAP 7-year value \citep{WMAP7}
and $h = 0.742 \pm 0.036$ from \cite{Riess09}.

The overdensity is often quoted as a factor $\Delta_{vir}$ 
above the mean background density $\rho_m = \Omega_m \rho_{crit}$:
\begin{equation}
  \rho_{halo}
= \Delta_c \rho_{crit}
  = \Delta_{vir} \rho_m
\end{equation}
With $\Delta_{c} = \Delta_{vir} \Omega_m$, 
$\Delta_c = 101.1$ corresponds to $\Delta_{vir} = 337$ for $\Omega_m = 0.3$.
This value is cited by e.g., \cite{Bullock01} and \cite{Graham06II}.
Using the \cite{Eke98} relation
and the \cite{Spergel03} first-year WMAP value of $\Omega_m = 0.268$,
\citet[among others]{Merritt06} give $\Delta_{vir} = 368$.

To facilitate comparison among current and future investigations,
we propose that a value of $\Delta_c = 100$ be adopted for present-day halos.
This corresponds to $\Delta_{vir} = 333$ given $\Omega_m = 0.3$,
or the nice round number $\Delta_{vir} = 400$ given $\Omega_m = 0.25$.
We also note that for $\Omega_m = 0.25$,
the \cite{BryanNorman98} expression yields 
$\Delta_c \approx 94$, $\Delta_{vir} \approx 376$.

While results from simulations are most often reported for present-day halos,
Nature provides us observers with images of clusters as they were in the past.
Thus in the expressions above, we should replace the present day values of
$\Omega_{m,0}$ and $\Omega_{\Lambda,0}$
(here ``$0$'' subscripts have been added for clarity) with:  
\\
\begin{equation}
  \Omega_m(z) = \frac{1}{1 + (\Omega_{\Lambda,0} / \Omega_{m,0}) (1 + z)^{-3}}
\end{equation}
\\
and $\Omega_\Lambda(z) = 1 - \Omega_m(z)$.
For the massive galaxy cluster A1689 at $z = 0.1862$ 
and adopting $\Omega_m = 0.3$, 
the widely used \cite{BryanNorman98} expression yields $\Delta_{c} = 116.6$
and the \cite{NakamuraSuto97} expression yields $\Delta_{c} = 115$.
The latter was adopted by
\citet[private communication]{Broadhurst05,Broadhurst05b}
so we adopt it as well for consistency in \cite{Coe10}.

In Figs.~\ref{OmOLz}, \ref{dc}, and \ref{dcvir},
we plot $\Omega_m(z)$, $\Delta_c(z)$, and $\Delta_{vir}(z)$.

\subsection{Conversion between overdensity values $\Delta_c$}

If the mass profile is well described by an NFW profile,
then it is straightforward to convert $c_{vir}$, $r_{vir}$, and $M_{vir}$
between different conventions of $\Delta_c$
(c.f., Fig.~\ref{fig:ccompareoverdens}).
Converting from $c_{200}$ ($\Delta_c = 200$) to $c_{115}$ ($\Delta_c = 115$),
for example, simply involves finding that value of $c_{115}$
which yields the same value of $\delta_c$ (Eq.~\ref{deltac}) as did $c_{200}$.
This can be accomplished by a simple root finding program,
but the relation is very linear as shown in Fig.~\ref{fig:cconv}.
Here we provide expressions
\begin{eqnarray} 
  \label{cconv3}
  c_{94}  & \approx & 1.328 c_{200} + 0.272\\
  c_{100} & \approx & 1.298 c_{200} + 0.246\\
  c_{115} & \approx & 1.232 c_{200} + 0.189
\end{eqnarray}
which are accurate to within 0.5\% for $2 < c_{200} < 25$.

These factors may be generalized:
\begin{equation}
  \label{cconveq}
  c_{vir} \approx a \thinspace c_{200} + b
\end{equation}
\vspace{-0.2in}
\begin{eqnarray}
  a & \approx & -1.119 \log_{10}\Delta_c + 3.537 \label{cconva}\\
  b & \approx & -0.967 \log_{10}\Delta_c + 2.181 \label{cconvb}
\end{eqnarray} 
to yield $c_{vir}$ within 1\% for $3 < c_{200}$ and $70 < \Delta_c < 140$.
These factors are plotted in Fig.~\ref{fig:Dc}.

An alternate expression
gives $c_{vir}$ as a function of $c_{200}$ and $\Delta_c$:
\\
\begin{equation}
  \label{cconvgen}
  c_{vir} = c_{200} + c_{200}^{0.9} 10^p
\end{equation}
\begin{equation}
  \label{cconvgenb}
  p = -(8.683 \times 10^{-5}) \Delta_c^{1.82}
\end{equation}
to within 1\% for $3 < c_{200} < 35$ and $85 < \Delta_c < 165$
($z < 1$ or so).

See also \citet[Appendix C]{HuKravtsov03}.

\subsection{Virial Mass}
\label{Mvir}

Virial mass (the mass within $r_{vir}$) is given by
\\
\begin{equation}
  \label{eq:Mvir}
  M_{vir} = \frac{4}{3} \pi r_{vir}^3 \Delta_c \rho_{crit}(z) = \frac{r_{vir}^3 \Delta_c  H^2(z)}{2G}
\end{equation}
In Fig.~\ref{Mrvir} we plot this simple relation between $r_{vir}$ and $M_{vir}$.

Quoted values for $M_{200}$ and $r_{200}$
can also be converted to $M_{vir}$ and $r_{vir}$
as a function of $c_{200}$ and $\Delta_c$
as plotted in Fig.~\ref{virrat}.
For a given NFW curve (with fixed $r_s$ and $\rho_s$),
$r_{vir}$ simply scales with $c_{vir}$ since $r_s$ stays fixed.
So $r_{vir} / r_{200} = c_{vir} / c_{200}$.
As for the $M_{vir}$ conversions, we solved for those numerically.

\section{Mass Profiles}
\label{massprofiles}

\subsection{Double Power Laws}
In dark matter simulations,
galaxy and cluster halos \cite{NFW96,NFW97} were all shown to have 
mass density profiles well approximated by the NFW profile:
\\
\begin{equation}
\label{NFWEQ}
  \rho(r) = \frac{\rho_s}{(r/r_s)(1+r/r_s)^2}.
\end{equation}
\\
This profile behaves as 
$\rho \propto r^{-1}$ in the core,
$\rho \propto r^{-2}$ at $r = r_s$, and steepens to
$\rho \propto r^{-3}$ in the outskirts.
The two fit parameters $\rho_s$ and $r_s$ were shown to be related
and a function of halo mass.
This ``universal'' profile is still a good approximation to today's simulated halos.
However the higher resolution does reveal subtle differences.

Deviations were sought for using a generalized version of the NFW profile 
\citep[][his Eq.~43; see also \citealp{Zhao96,Wyithe01}]{Hernquist90}:
\\
\begin{equation}
  \label{eqabc}
  \rho(r) = \frac{2^{(\beta-\gamma)/\alpha} ~ \rho_s}
  {(r/r_s)^\gamma [1+(r/r_s)^\alpha]^{(\beta-\gamma)/\alpha}}.
\end{equation}
\\
This profile behaves as 
$\rho \propto r^{-\gamma}$ in the core, and
$\rho \propto r^{-\beta}$ in the outskirts.
The rate of transition is governed by $\alpha$.
Where NFW found ($\alpha$, $\beta$, $\gamma$) = (1, 3, 1),
\cite{Moore99} instead found best fits of (1.5, 3, 1.5).
Importantly, the inner profile was steeper: $\gamma = 1.5$, $\rho \propto r^{-1.5}$.
There were many other attempts to accurately resolve and measure this inner slope,
including \cite{Diemand05} who found $\rho \propto r^{-1.2}$.

The fully generalized form in Equation \ref{eqabc} proves a bit too general 
with large degeneracies between the free parameters \citep{Klypin01}.
Thus, in their efforts to determine the central slope $\gamma$,
authors often use one of two constrained versions of Equation \ref{eqabc},
either a ``generalized NFW'' profile with
($\alpha$, $\beta$, $\gamma$) = (1, 3, $\gamma$):
\\
\begin{equation}
  \label{genNFW}
  \rho(r) = \frac{2^{3-\gamma} \rho_s}{(r/r_s)^\gamma [1+(r/r_s)]^{(3-\gamma)}},
\end{equation}
\\
or what we might call a ``generalized Moore'' profile\footnote{This latter form
is also often referred to as a ``generalized NFW'' profile,
although strictly speaking it can only exactly reproduce the Moore profile
and not that of NFW.}
with
($\alpha$, $\beta$, $\gamma$) = ($3-\gamma$, 3, $\gamma$):
\\
\begin{equation}
  \label{genMoore}
  \rho(r) = \frac{2 \rho_s}{(r/r_s)^\gamma [1+(r/r_s)^{3-\gamma}]}.
\end{equation}
\\

Meanwhile, \cite{DehnenMcLaughlin05} found\\
($\alpha$, $\beta$, $\gamma$) = (4/9, 31/9, 7/9):
\begin{equation}
  \label{DehnenMcLaughlin2}
  \rho(r) = \frac{2 \rho_s^6}{(r/r_s)^{7/9} [1+(r/r_s)^{4/9}]^6}.
\end{equation}
\\
and when accounting for anisotropy, the more general\\
\{$\alpha$, $\beta$, $\gamma$\} = 
\{$(3-\gamma)/5$, $(18-\gamma)/5$, $\gamma$\}:
\begin{equation}
  \label{DehnenMcLaughlin3}
  \rho(r) = \frac{2 \rho_s^6}{(r/r_s)^{\gamma} [1+(r/r_s)^{(3-\gamma)/5}]^6}.
\end{equation}

\subsection{Continuously Varying Power Laws}

The original NFW proponents proposed a new profile
which gradually flattens all the way toward the center \citep{Navarro04}.
This profile was found \citep{Navarro04,Merritt05,Merritt06}
to yield better fits to a wide range of simulated dark matter halos
than did the generalized NFW profile (Eq.~\ref{genNFW}),
which has an equal number (3) of free parameters, including the central slope.
Inner slopes as steep as $\rho(r) \propto r^{-1.2}$ 
are clearly ruled out by recent simulations \citep{Navarro10}.

The new \cite{Navarro04} fitting form was quickly recognized \citep{Merritt05}
as the \cite{Sersic68} profile
generally applied to fitting the light distributions of elliptical galaxies.
The implications are intriguing: 
that the collapse of massive bodies,
be they luminous or dark matter,
may lead to similar profiles.

However to be precise, 
\cite{Navarro04} fit a S\'ersic-like profile to 3-D density distributions,
where the S\'ersic profile was fit to 2-D surface density distributions (of light).
\cite{Einasto65} was first to use such a density law to describe a 3-D distribution,
namely the spatial distribution of old stars within the Milky Way.

Today we distinguish between the ``Einasto'' and ``S\'ersic'' mass profiles.
The former is fit to 3-D mass density $\rho(r)$
while the latter is fit to 2-D projected mass distributions $\Sigma(R)$.
Projected and deprojected approximations to 
the Einasto and S\'ersic profiles, respectively,
have also been derived (see Table \ref{tab:projects}).

The \cite{Sersic68} profile is given by:
\\
\begin{equation}
\label{SersicEq}
  \Sigma(R) = \Sigma_e \exp\left\{
    -b_n \left[ \left( \frac{R}{R_e} \right)^{1/n} - 1 \right]
  \right\},
\end{equation}
\\
There are three free parameters: $\Sigma_e$, $R_e$, and $n$,
with $b_n$ being a function of $n$ (given below) such that
half the mass is contained within $R_e$.
Note that the total mass of a S\'ersic profile is finite,
unlike that for an NFW profile.
A 3-D deprojected approximation is given by \cite{PrugnielSimien97}.

The Einasto mass profile is a similar function but of 3-D mass density $\rho(r)$:
\\
\begin{equation}
  \label{Einasto}
  \rho(r) = \rho_{-2} \exp \left( -\frac{2}{\alpha} \left[ \left( \frac{r}{r_{-2}} \right)^\alpha - 1 \right]  \right)
\end{equation}
\\
where $\rho_{-2}$ and $r_{-2}$ are the density and radius 
at which $\rho(r) \propto r^{-2}$.
The concentration is defined as $c_{vir} = r_{vir} / r_{-2}$.
\cite{Navarro10} found $\alpha \approx 0.17$ for galaxy-sized halos 
in the Aquarius simulation.
\cite{Gao08} concur and found $\alpha$ increases to $\sim 0.3$ 
for the most massive clusters in the Millennium simulation.
\cite{Duffy08} reduce the Einasto profile to two free parameter
by using the ``peak height'' $\alpha(\nu)$ relation from \cite{Gao08}.
A 2-D projected approximation of Einasto is given by \cite{DharWilliams10}.

\cite{Merritt05,Merritt06} experimented with both of these and other fits 
to 3-D mass density profiles of simulated halos.
In the latter paper, they compared the performance of 
various formulae fit to 10 simulated halos (6 cluster-sized and 4 galaxy-sized).
Their results for 3-parameter fits are reprinted here in 
Table \ref{tab:Merritt06profilefits}.
The four tested profiles yielded similar results.
Einasto performed a bit better across the board.
Prugniel-Simien (the deprojected S\'ersic profile) performed a bit better for clusters.
Dehnen-McLaughlin performed a bit better for galaxies.
And the generalized NFW profile (Eq.~\ref{genNFW}) was not far behind.

Given the small number of halos tested, and the similarity of the performances,
all of these fitting formulae might still be considered reasonable choices.
However the Einasto profile has become especially popular
\cite[e.g.,][]{HayashiWhite08,Gao08,Duffy08,Navarro10}.

Recently a new fitting formula was proposed by \cite{Stadel09}.
It yields superior fits to the two high resolution halos tested
(Via Lactea 2 and GHALO):
\\
\begin{equation}
  \label{StadelMoore}
  \rho(r) = \rho_0 e^{ -\lambda \left[ \ln (1 + r / R_\lambda)\right]^2}
\end{equation}
\\
with $\lambda \simeq 0.1$.

\begin{deluxetable}{ccc}
\tablecaption{\label{tab:projects}
S\'ersic-like profiles of 2-D and 3-D density}
\tablewidth{0pt}
\tablehead{
\colhead{3-D $\rho(r)$} &
\colhead{} &
\colhead{2-D $\Sigma(R)$}
}
\startdata
\cite{Einasto65} & $\underrightarrow{\rm projected}$ & \cite{DharWilliams10}
\vspace{0.1in}\\
\cite{PrugnielSimien97} & $\underleftarrow{\rm deprojected}$ & \cite{Sersic68}\\ 
\vspace{-0.1in}
\enddata
\end{deluxetable}

\begin{deluxetable}{lcc}
\tablecaption{\label{tab:Merritt06profilefits}
3-parameter fits: deviations from halo profiles
measured in \citet[their Table 4]{Merritt06}}
\tablewidth{0pt}
\tablehead{
\colhead{Model} &
\colhead{6 clusters} &
\colhead{4 galaxies}
}
\startdata
Einasto 
& 0.028 & 0.026\\
Prugniel-Simien\tablenotemark{a}
& 0.025 & 0.030\\
generalized NFW
& 0.032 & 0.028\\
Dehnen-McLaughlin
& 0.034 & 0.023\\
\vspace{-0.1in}
\enddata
\tablenotetext{a}{deprojected S\'ersic}
\end{deluxetable}

\subsection{Power Law Behaviors}

Power law slopes for the above fitting formulae as a function of $r$
are plotted in Figs.~\ref{powerlaws}, \ref{powerlaws_SMDM}, \ref{powerlaws_DM}.
The slopes were calculated numerically as $d \ln \rho / d \ln r$.
For fun, we note this is equivalent to $(d\rho / d r) (r / \rho)$.
The radii $r$ are given in units of $r_{-2}$ at which $\rho \propto r^{-2}$.

We see that in principle, 
we should be able to distinguish between these various profiles
in both observed and simulated halos
given sufficient resolution at a large enough range of radii.
Such clear determinations have so far eluded us.

In Fig.~\ref{NFWpowerlaws}, 
we plot power law slopes for $\rho(r)$, $\kappa(R)$, and $M(<R)$ for the NFW profile.

\newpage  

\subsection{Profile Details}

Here we provide useful expressions derived from the NFW and S\'ersic profiles.

\subsubsection{NFW Profile}
\label{NFW}

Simulated galaxy and cluster halos \cite{NFW96,NFW97}
were shown to all have mass density profiles well approximated by the NFW profile:
\\
\begin{equation}
\label{NFWEQ}
  \rho(r) = \frac{\rho_s}{(r/r_s)(1+r/r_s)^2}.
\end{equation}
The two fit parameters $\rho_s$ and $r_s$ were shown to be related
and a function of halo mass, as we discuss below.
But the parameter making all the buzz these days is the central mass concentration:
\\
\begin{equation}
  \label{c}
  c_{vir} = r_{vir} / r_s,
\end{equation}
where $r_{vir}$ is the virial radius of the mass halo.
As discussed above, the virial radius is estimated as
that which contains an average density $\Delta_c \rho_{crit}$,
for a total virial mass of
\\
\begin{equation}
\label{Mvir}
  M_{vir} = \frac{4}{3} \pi \Delta_{c} \rho_{crit} r_{vir}^3.
\end{equation}

For an NFW halo, 
the mass within a sphere with radius $r = x r_s$
can be found by simply integrating the NFW profile (Eq.~\ref{NFWEQ}):
\\
\begin{eqnarray}
  M(r) & = & 4 \pi r_s^3 \int_0^x dx^\prime x^{\prime 2} \frac{\rho_s}{x^\prime (1+x^\prime)^2}\\
  & = & 4 \pi \rho_s r_s^3 \left( \ln(1+x) - \frac{x}{1+x} \right) 
\label{MNFW}
\end{eqnarray}
\\
Combining Eqs.~\ref{c}, \ref{Mvir}, and \ref{MNFW}, 
we find that the concentration parameter $c$ can be obtained from
the following expression, as given in \cite{NFW96}:
\\
\begin{equation}
  \label{deltac}
  \frac{\rho_s}{\rho_{crit}} \equiv \delta_c = 
  \frac{\Delta_c}{3} ~ \frac{c^3}{\ln(1+c) - c/(1+c)}.
\end{equation}

To fit the NFW profile to our gravitational lensing mass maps
which measure projected surface density,
we integrate the NFW profile along the line of sight
\cite[e.g.,][]{GolseKneib02}
to find the projected surface density:
\\
\begin{equation}
  \Sigma(R) = 2 \rho_s r_s F(X)
\end{equation}
\\
with $R = X r_s$ and
\\
\begin{eqnarray}
  \label{FX}
  F(X) = \left\{ 
    \begin{array}{ll}
      \displaystyle\frac{1}{X^{2}-1}
      \left(1-\frac{1}{\sqrt{1-X^{2}}}\cosh^{-1}\frac{1}{X}\right) 
      & (X<1)\\
      \displaystyle\frac{1}{3} 
      ~~~(X=1)\\
      \displaystyle\frac{1}{X^{2}-1}
      \left(1-\frac{1}{\sqrt{X^{2}-1}}\cos^{-1}\frac{1}{X}\right) 
      & (X>1)
    \end{array}\right.
\end{eqnarray}
\\
Integrating once more over the area within $R$, 
we find the total mass within a {\it cylinder} of radius $R$
\\
\begin{equation}
  \label{NFWMcyl}
  M(R) = 4 \pi  r_s^3 \rho_s G(X)
\end{equation}
\\
with
\\
\begin{eqnarray}
  \label{NFWGcyl}
  G(X) = \ln \frac{X}{2} + \left\{ 
    \begin{array}{ll}
      \displaystyle\frac{1}{\sqrt{1-X^{2}}}\cosh^{-1}\frac{1}{X}
      & (X<1)\vspace{0.1in}\\
      \displaystyle 1
      \vspace{0.1in}
      & (X=1)\\
      \displaystyle\frac{1}{\sqrt{X^{2}-1}}\cos^{-1}\frac{1}{X}
      & (X>1)
    \end{array}\right.
\end{eqnarray}
\\
This should not be confused with Eq.~\ref{MNFW} 
which gives mass within a {\it sphere} of radius $r$.

From this we can obtain the shear due to an NFW mass profile:
$\gamma(R) = \bar\kappa(R) - \kappa(R)$:
\\
\begin{equation}
  \gamma(R) = 2 \kappa_s \left( \frac{2 G(X)}{X^2} - F(X) \right).
\end{equation}
\\
The quantity measured in weak lensing studies is the reduced shear:
\\
\begin{equation}
  g = \frac{(D_{LS}/D_S) \gamma}{1 - (D_{LS}/D_S) \kappa},
\end{equation}
\\
where we have finally given the redshift dependence.
(All previous expressions were given for a fiducial lensed source at $z_s=\infty$.)

\subsubsection{S\'ersic Profile}
\label{Sersic}

We now give the \cite{Sersic68} profile and quantities derived from it
\citep[e.g.,][]{GrahamDriver05,TerzicGraham05}.
Note that the S\'ersic profile is commonly used to describe 
the (projected 2-D) light profiles of elliptical galaxies.
Here instead it will be discussed as describing projected mass profiles.

The \cite{Sersic68} profile is given by:
\begin{equation}
\label{SersicEq}
  \Sigma(R) = \Sigma_e \exp\left\{
    -b_n \left[ \left( \frac{R}{R_e} \right)^{1/n} - 1 \right]
  \right\},
\end{equation}
There are three free parameters: $\Sigma_e$, $R_e$, and $n$,
with $b_n$ being a function of $n$ (given below) such that
half the mass is contained within $R_e$.
(Note that the total mass of a S\'ersic profile is finite,
unlike that for an NFW profile.)
The total projected mass within a radius $R$ is given as:
\begin{equation}
  \label{MSersic}
  M(R) = 2 \pi \Sigma_e R_e^2 n e^{b_n} b_n^{-2n} \gammainc(2n, x)
\end{equation}
where
\begin{equation}
  \label{bn}
 x = b_n (R / R_e) ^ {1/n},
\end{equation}
%
and $\gammainc(a,x) = \int_0^x dt e^{-t} t^{a-1}$
is the incomplete gamma function
(with the ``hat'' used to distinguish $\gammainc$ 
from the lensing shear $\gamma$).
Thus to satisfy $M(R_e) = \onehalf M(R = \infty)$,
$b_n$ must obey:
\begin{equation}
  \label{bn}
  \Gamma(2n) = 2 \gammainc(2n, b_n)
\end{equation}
where $\Gamma(a) = \gammainc(a,\infty)$ is the complete gamma function.
In SciPy's ``special'' package, we find a routine to quickly calculate 
$b_n = {\tt gammaincinv(2*n, 0.5)}$.
An approximation may also be used \citep{PrugnielSimien97}:
\begin{equation}
  \label{bnapprox}
  b_n \approx 2n - 1/3 + 0.009876/n
\end{equation}

Lensing properties of the S\'ersic profile
have been derived and explored in \cite{Cardone04} and \cite{Eliasdottir07}.
Of special interest here is the weak shear $\gamma = \bar\kappa - \kappa$.
The average $\kappa$ within $R$ can be derived straightforwardly
from the above expression for $M(R)$:
\\
\begin{equation}
  \label{kavgSersic}
  \bar\kappa(R) = \frac{M(R)}{\pi R^2 \Sigma_{crit}}
    = 2 \kappa_e n e^{b_n} x^{-2n} \gammainc(2n, x)
\end{equation}
where we have introduced $\kappa_e = \Sigma_e / \Sigma_{crit}$.
Meanwhile, $\kappa(R) = \Sigma(R) / \Sigma_{crit}$ can be rewritten as:
\begin{equation}
  \label{kSersic}
  \kappa(R) = \kappa_e e^{(b_n - x)}
\end{equation}
Thus we find $\gamma(R) = \bar\kappa(R) - \kappa(R)$:
\begin{equation}
  \label{ySersic}
  \gamma(R) = \kappa_e e^{b_n} \left( 2 n x^{-2n} \gammainc(2n, x) - e^{-x} \right),
\end{equation}
with the reduced shear given as
$g = (\gamma D_{LS}/D_S) / (1 - \kappa D_{LS}/D_S)$.


There are fewer published fits of S\'ersic profiles to simulated cluster halos.
We do note that \cite{Merritt05} found $n = 2.38 \pm 0.25$
for their cluster sample.



\vspace{0.3in}
\acknowledgements{We thank Angelo Neto for useful conversations 
about the Millennium simulation and their study of halo profiles.
This work was carried out at Jet Propulsion Laboratory,
California Institute of Technology, under a contract with NASA.}

\vfill

\begin{figure*}
\plottwo{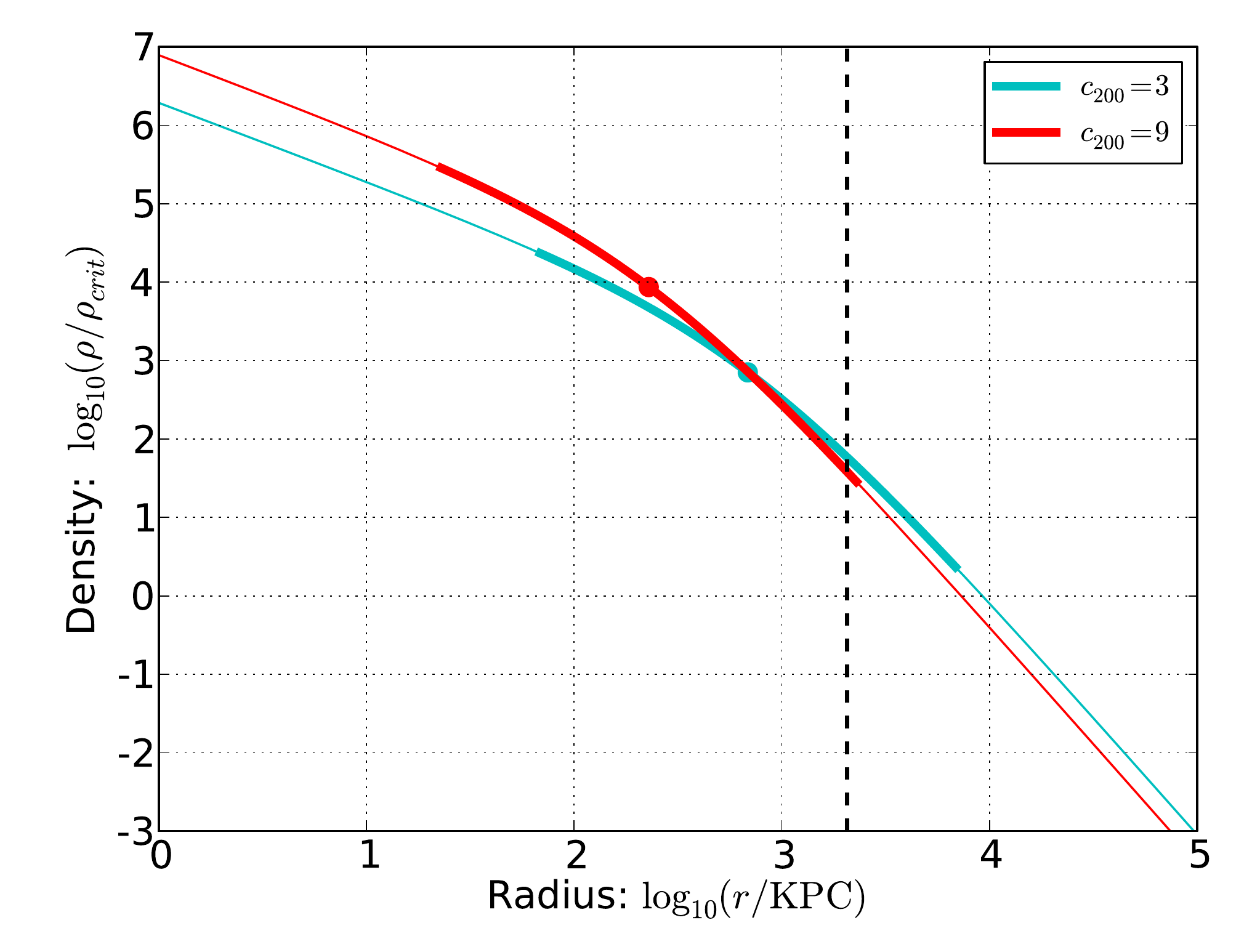}{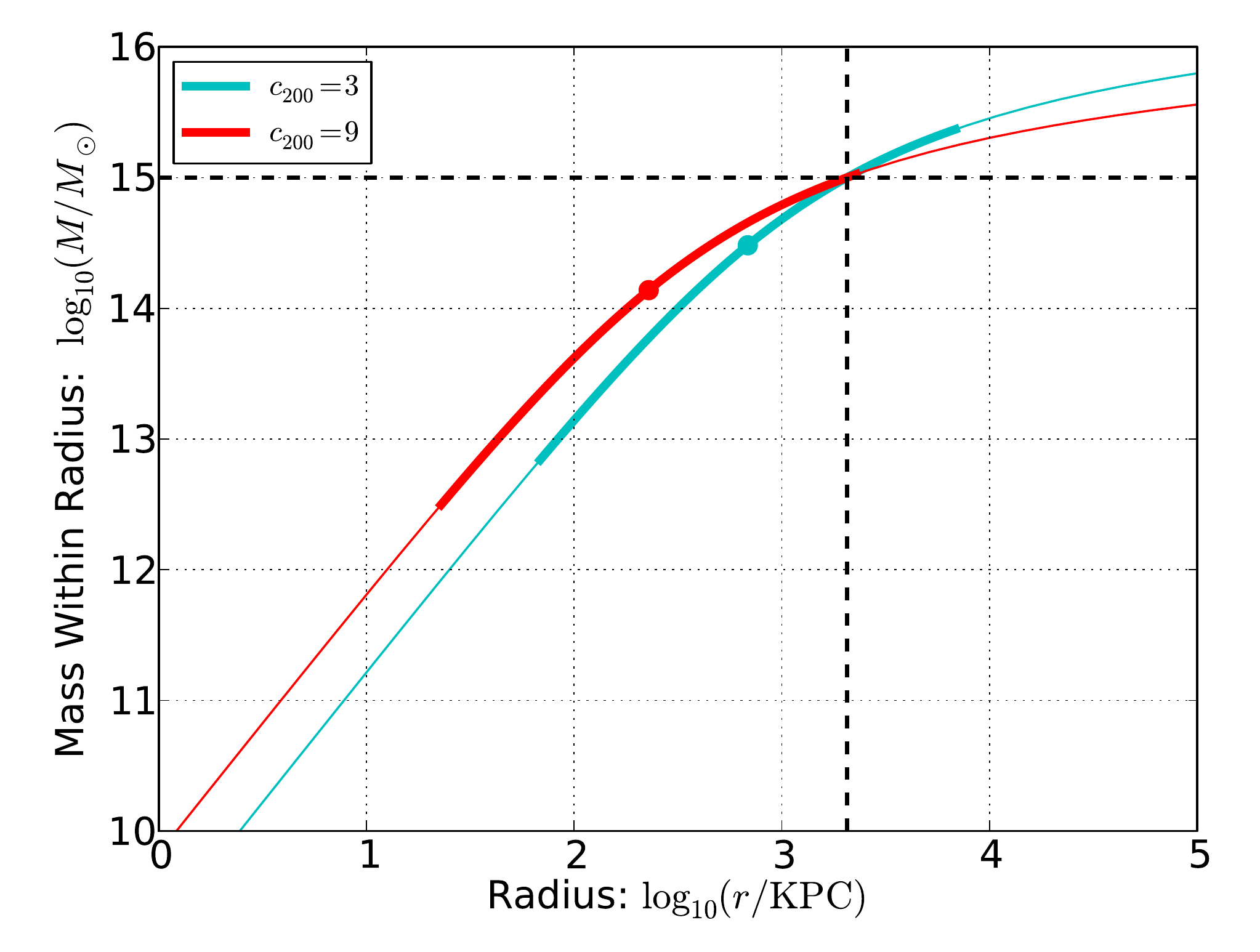}
\caption[]{\label{fig:ccompare}%
{\bf For fixed $\mathbf{M_{200} = 10^{15} M_\odot}$},
we show what an ``over-concentrated'' looks like:
$c_{200} = 9$ versus the expected concentration ($c_{200} = 3$).
This is roughly the case for A1689 \citep{Coe10}.
Dots mark $r_s$ and dashed lines mark $r_{vir}$ and $M_{vir}$.
Thicker lines are used for two decades in radius about $r_s$
to emphasize the shifting of the curves.
}\end{figure*}


\begin{figure*}
\plottwo{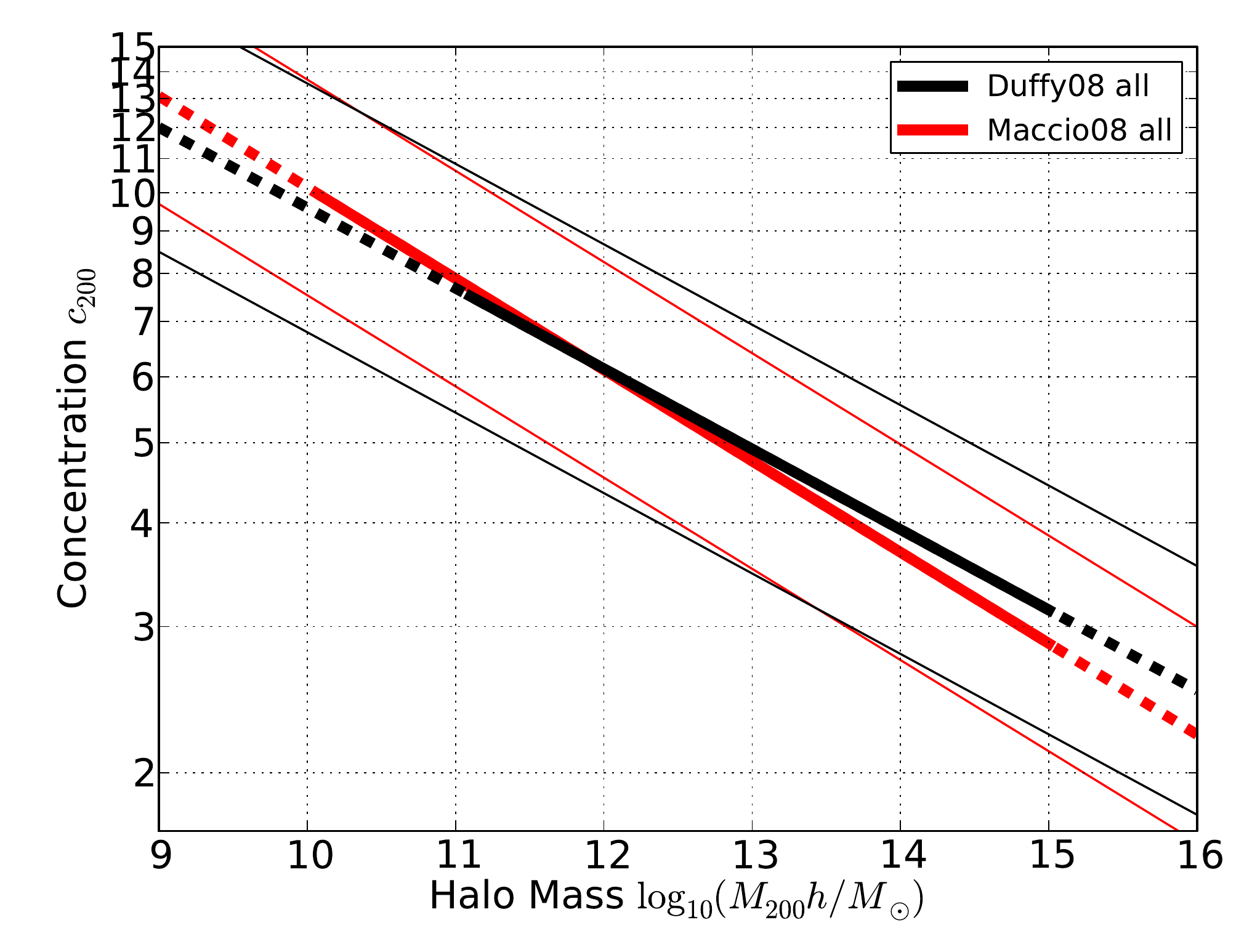}{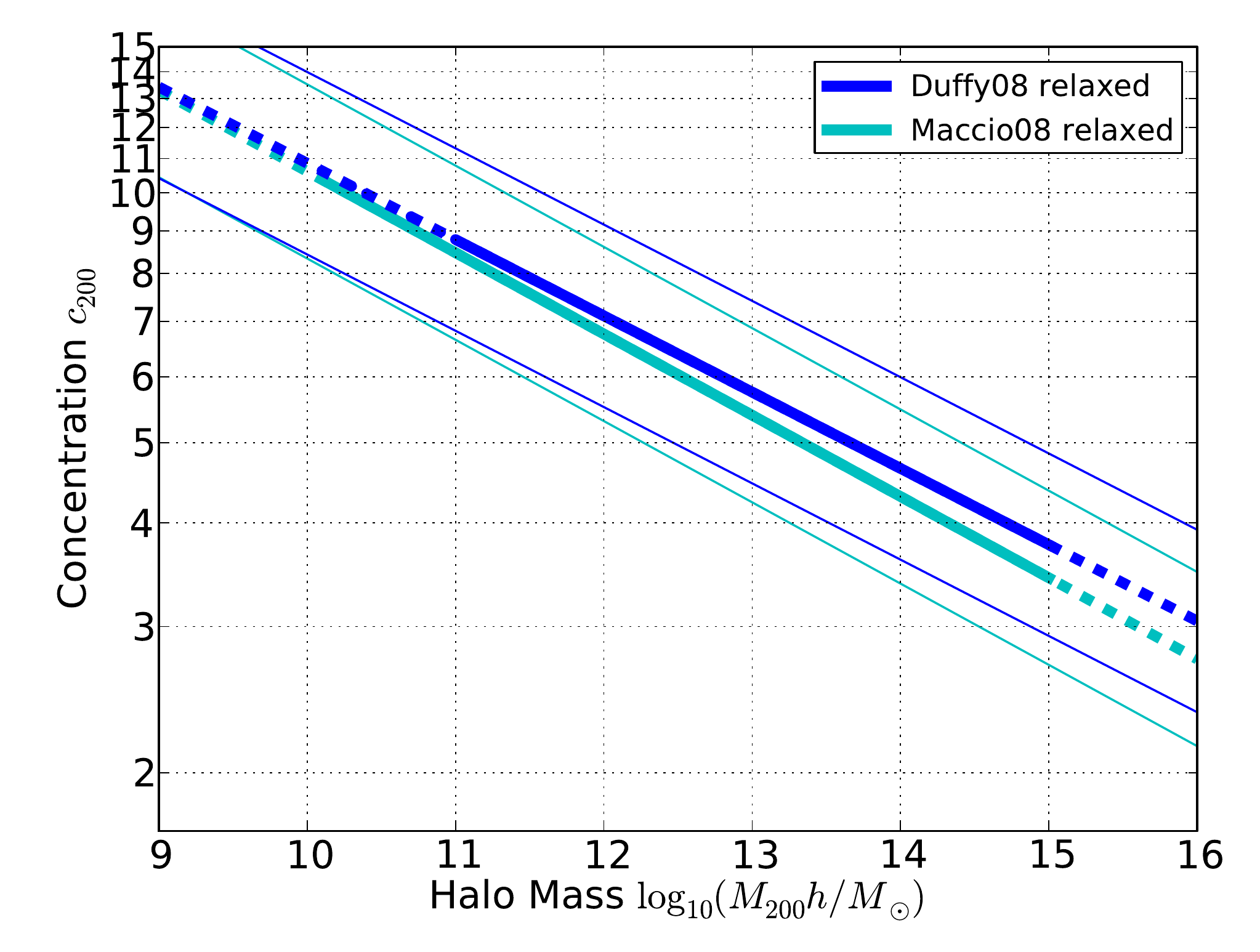}
\caption[]{\label{cMDuffyMaccio}%
Expected NFW concentration $c_{200}$ and 1-$\sigma$ scatter
as a function of halo mass $M_{200}$
for all clusters (left) and relaxed clusters (right).
}\end{figure*}

\begin{figure*}
\plottwo{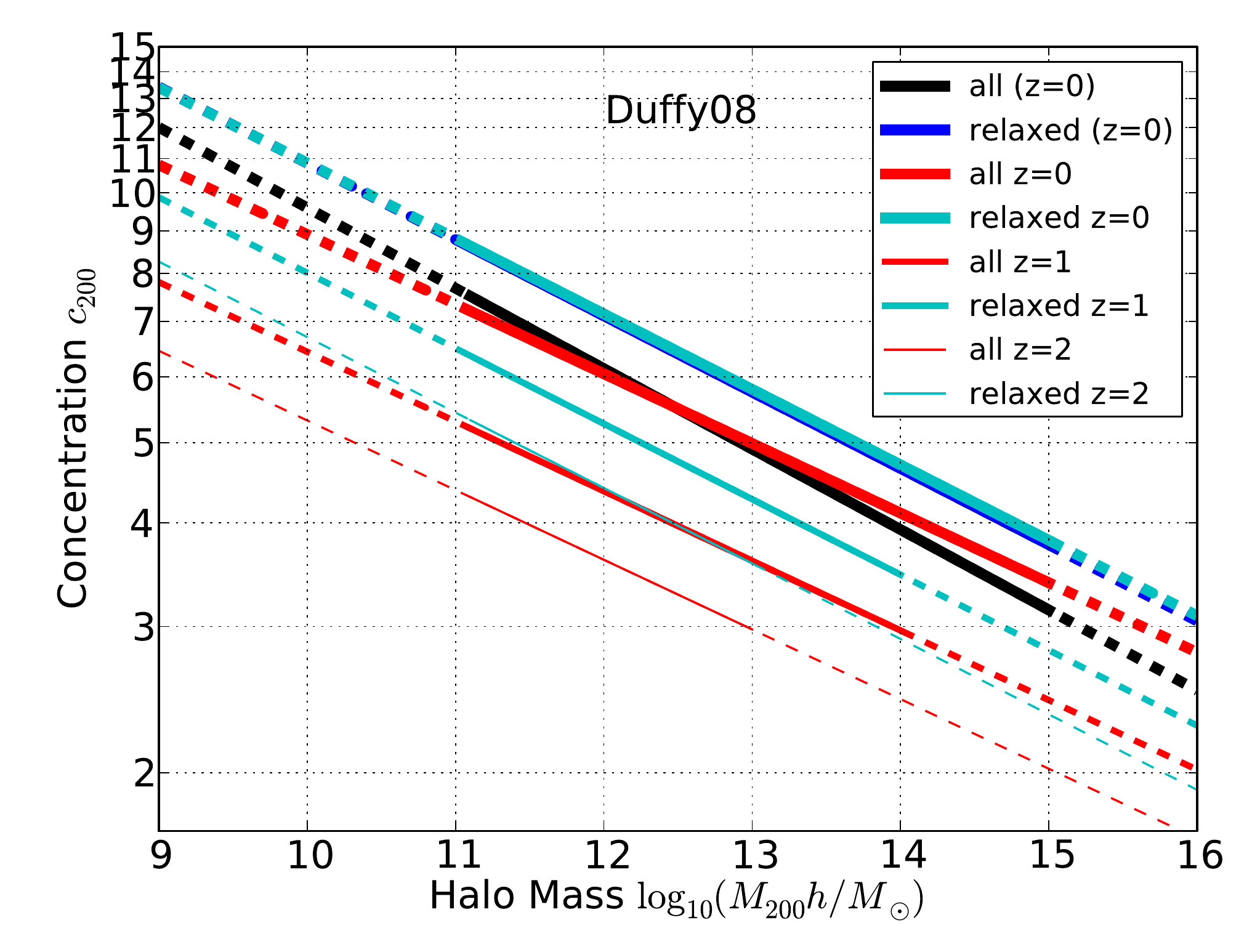}{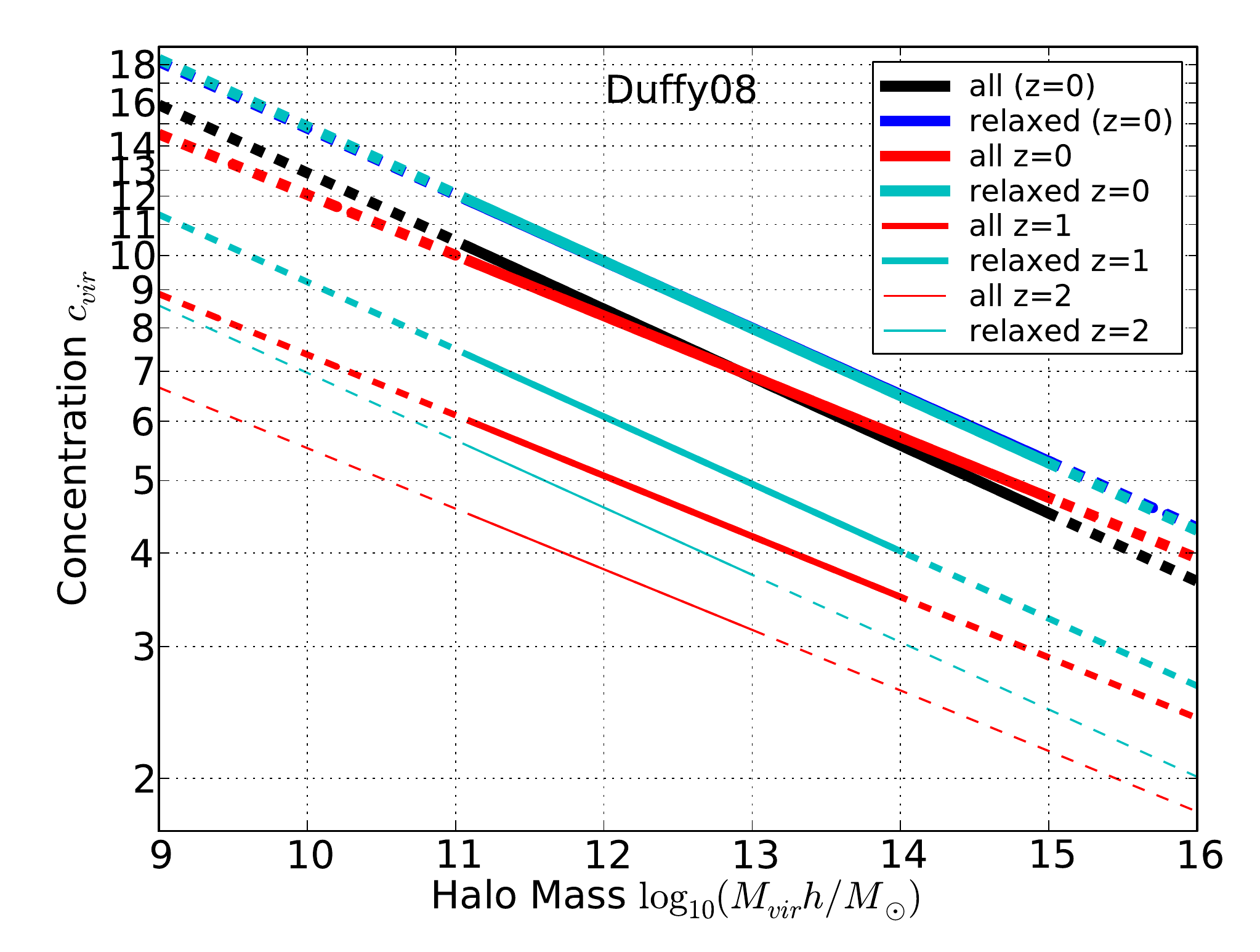}
\caption[]{\label{cMzM}%
Expected NFW concentration $c(M)$ for $z=0$ halos
and $c(M,z)$ for $z=0,1,2$ halos from \cite{Duffy08}.
}\end{figure*}

\begin{figure*}
\plottwo{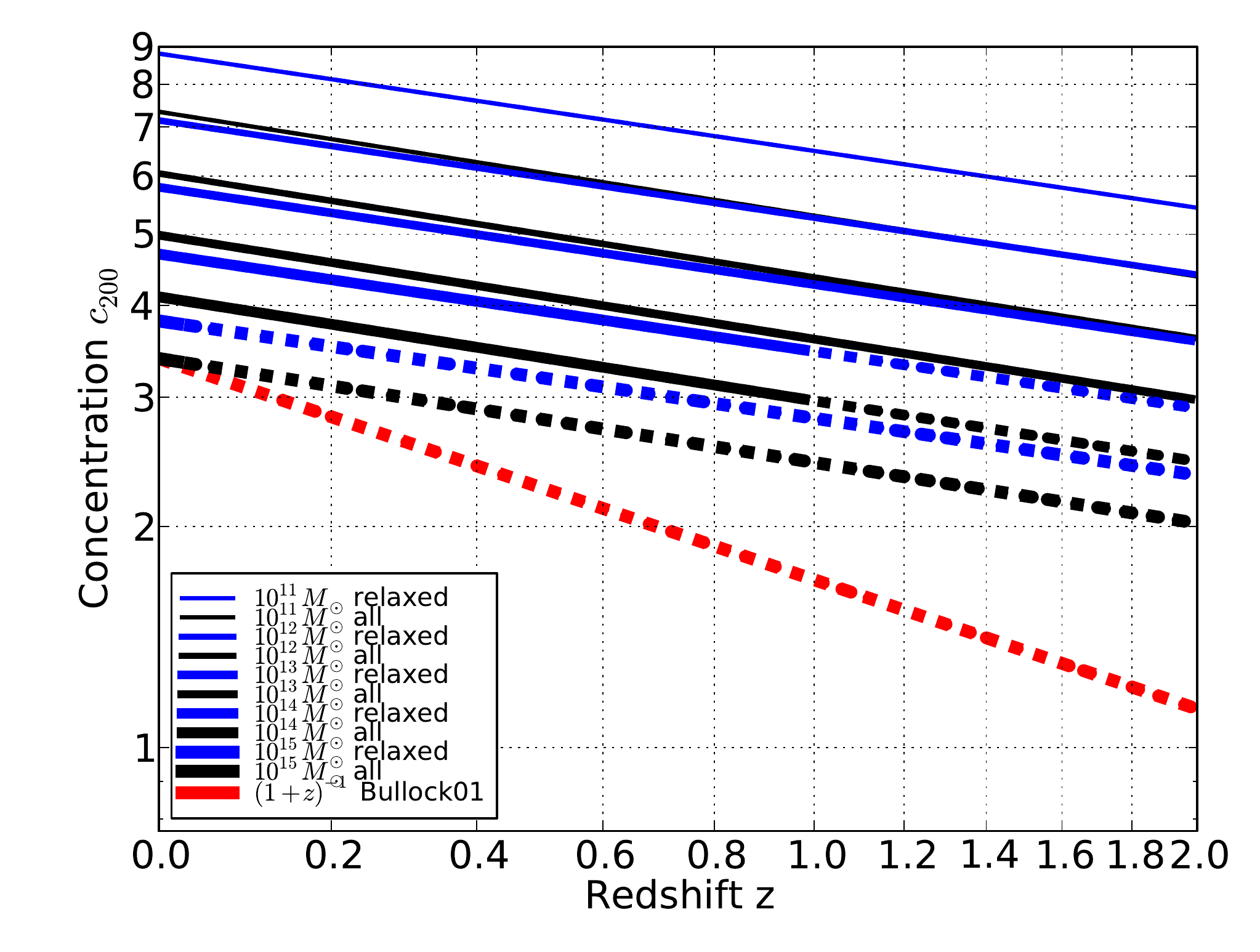}{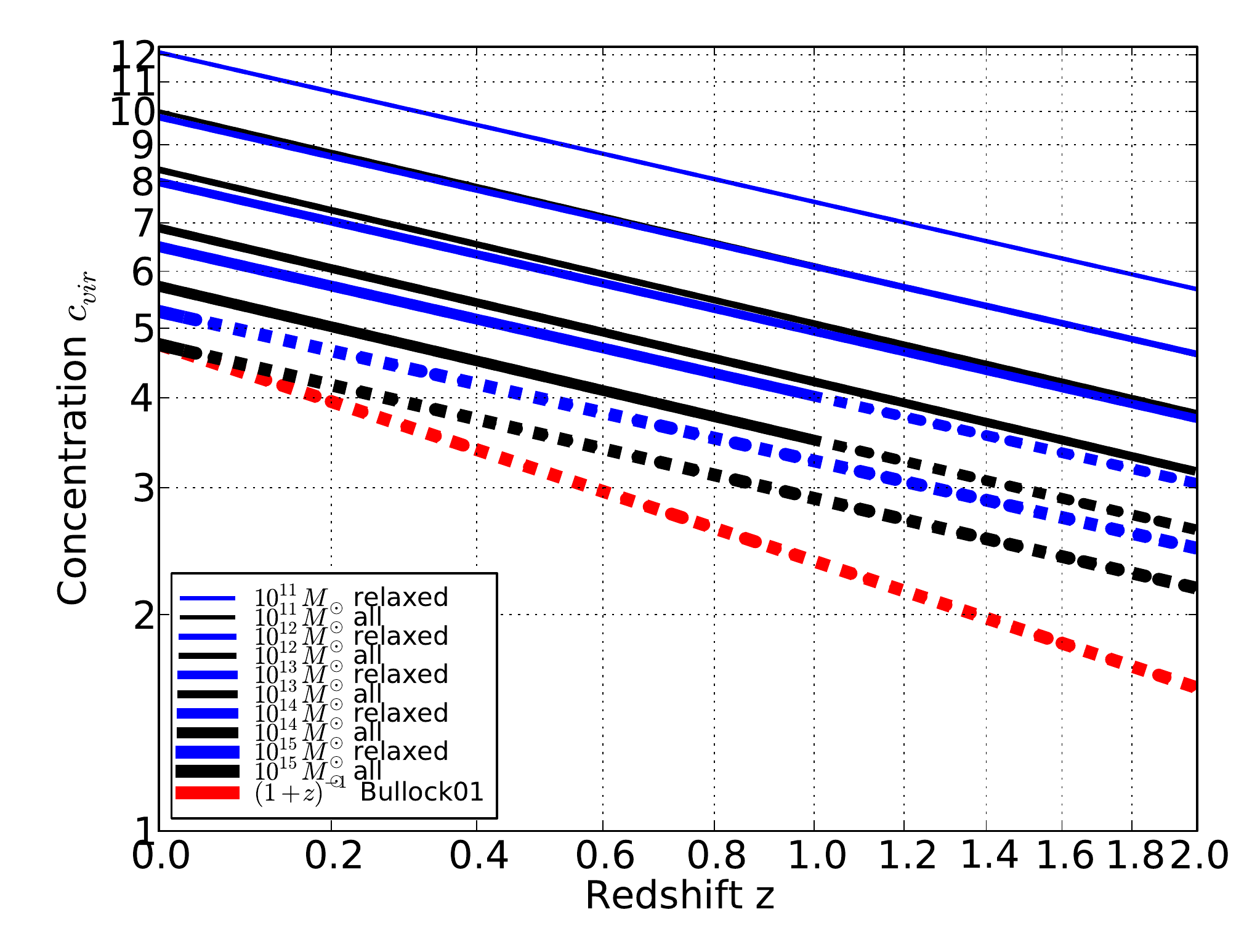}
\caption[]{\label{cMzz}%
Expected NFW concentrations as a function of redshift for various mass halos
from \cite{Duffy08}
(see Table \ref{tab:cMz}).
Also plotted is the $c \propto (1+z)^{-1}$ slope expected from \cite{Bullock01}.
The $x$ axes are plotted on scales of $\log(1+z)$.
}\end{figure*}

\begin{figure*}
\plottwo{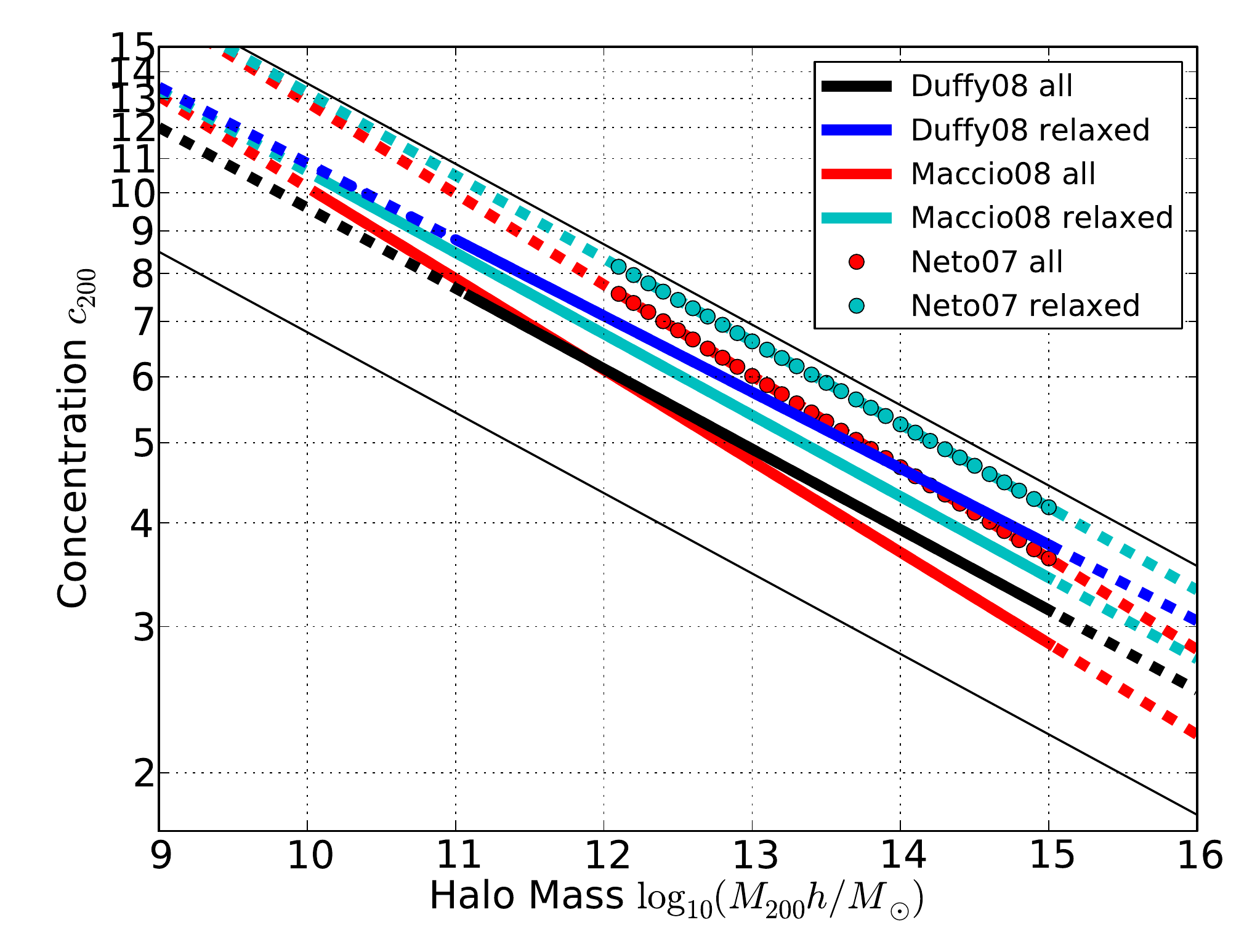}{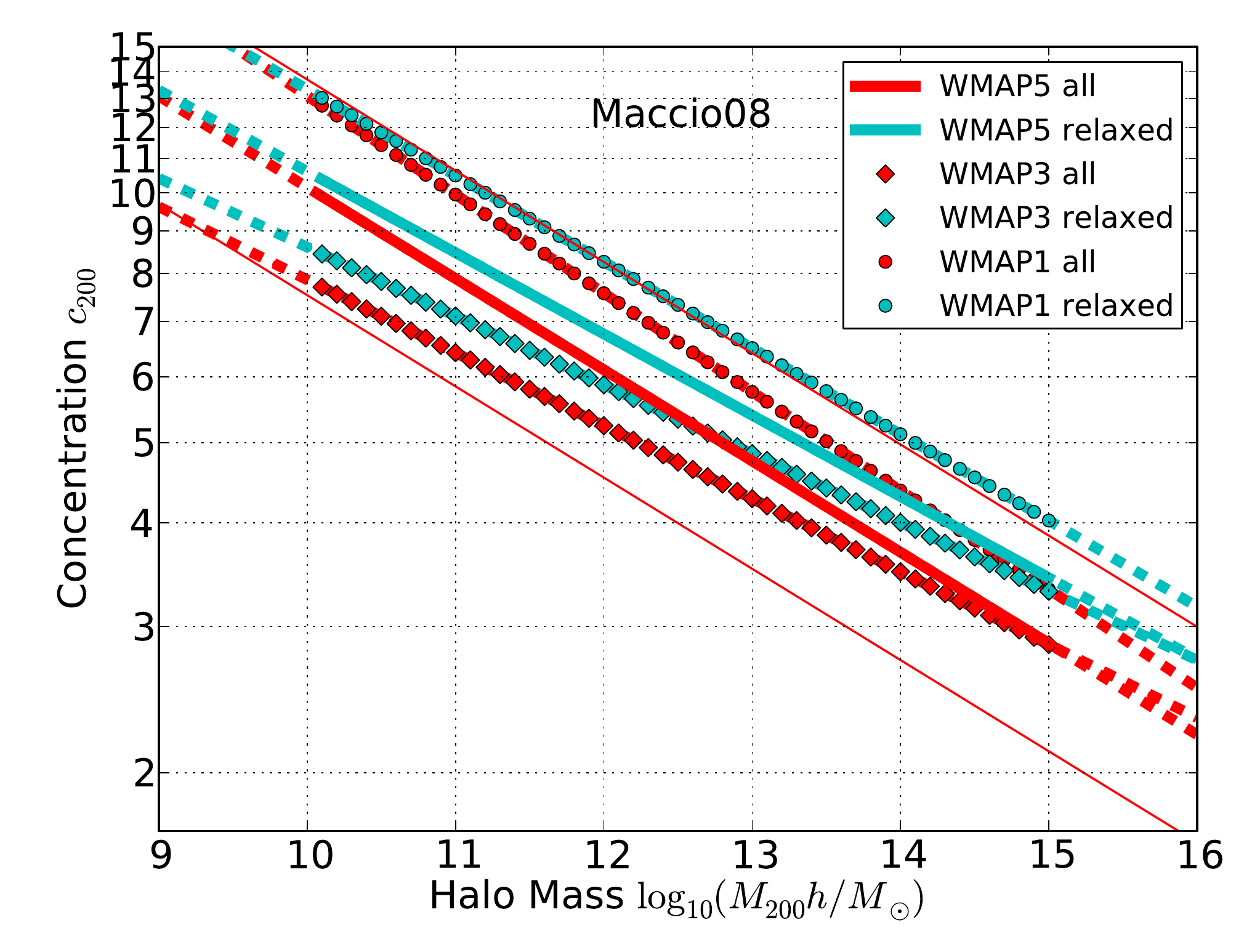}
\caption[]{\label{cMcosmo}%
The effect of cosmology.
Left: the Millennium simulation (Neto07) 
used WMAP1 with a higher $\sigma_8$ resulting in higher concentrations.
Right: 3 different simulations with different cosmologies used.
}\end{figure*}

\begin{figure*}
\plottwo{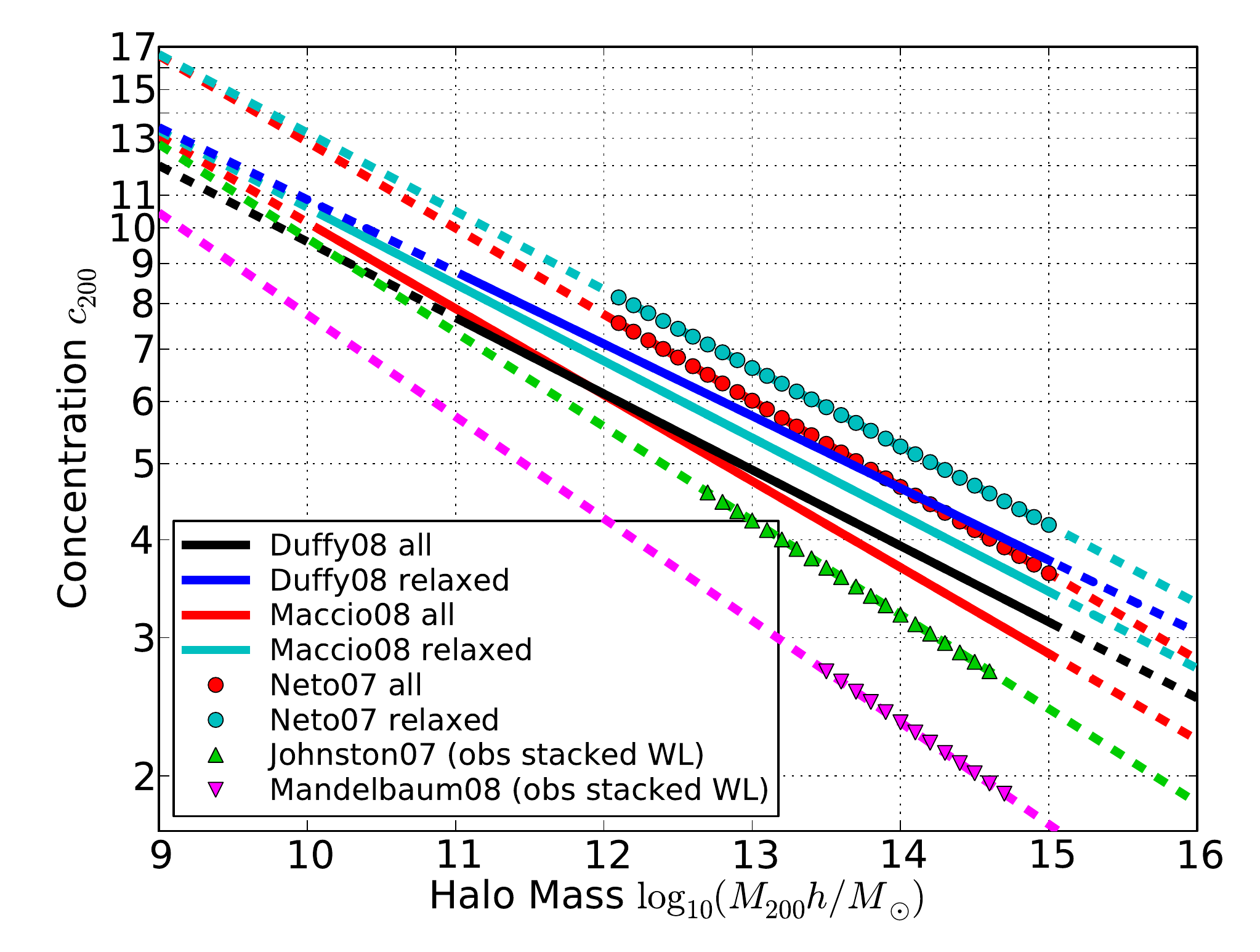}{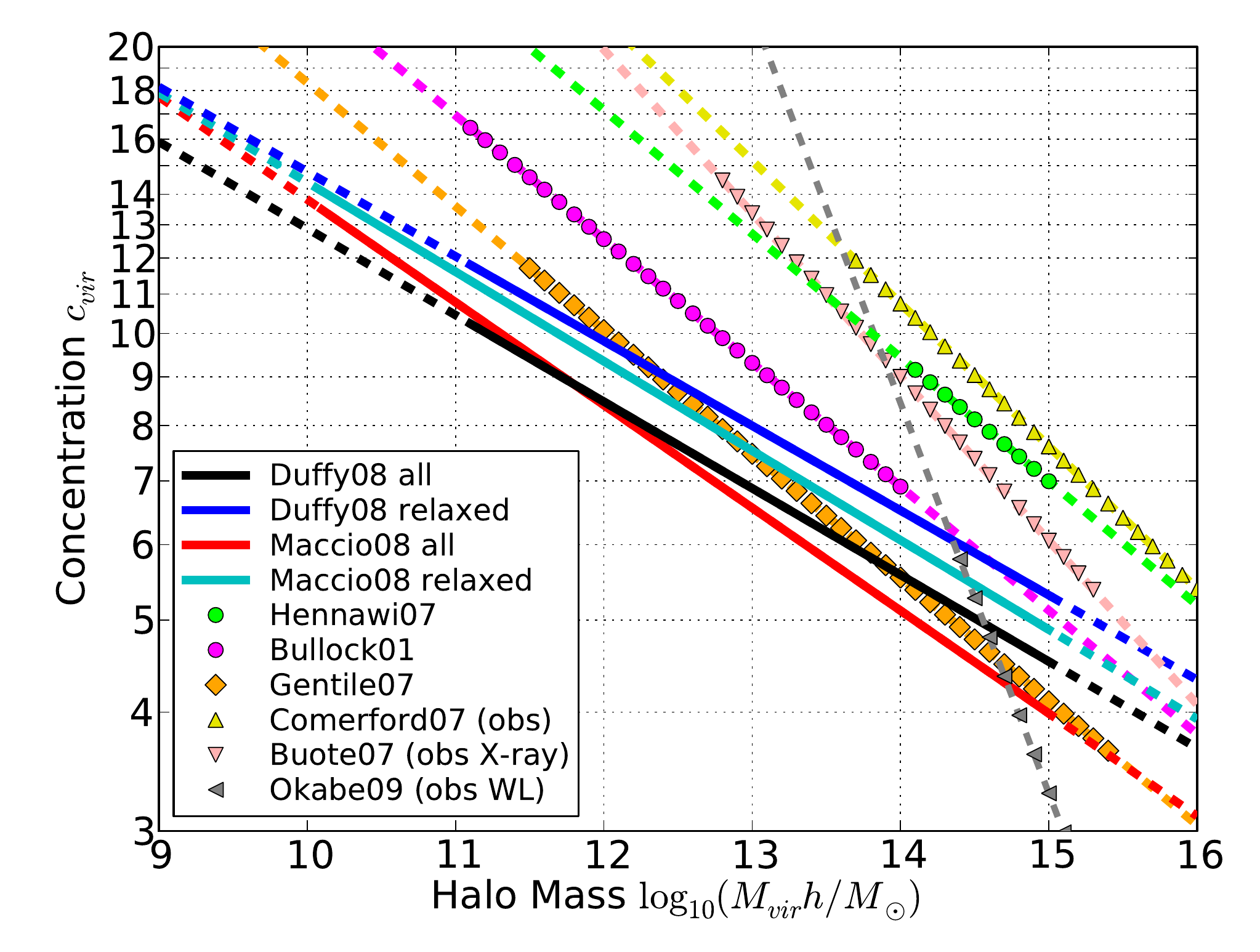}
\caption[]{\label{cMcompare}%
Comparing several studies, including both simulated and observed NFW $c(M)$.
}\end{figure*}

\begin{figure}
\plotone{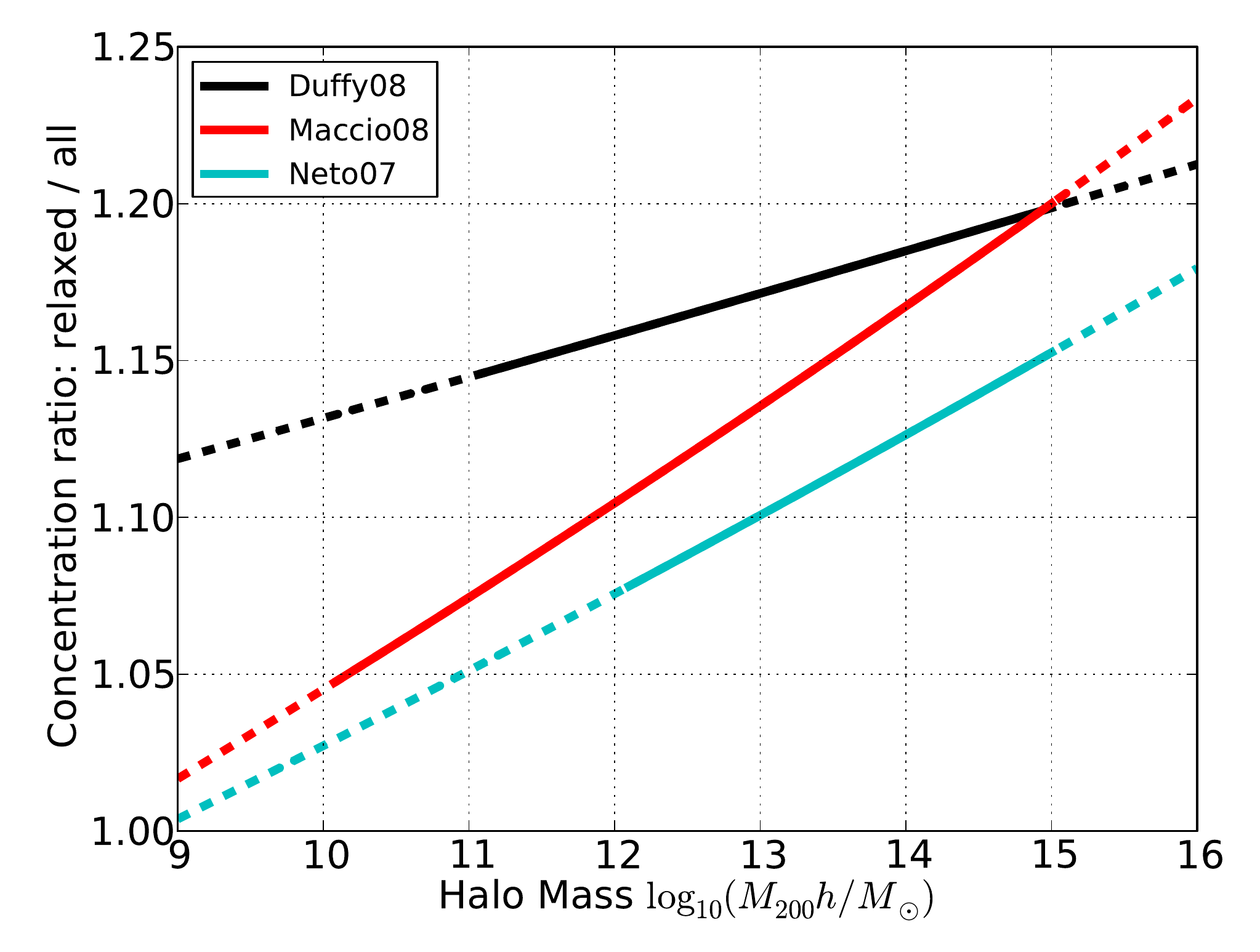}
\caption[]{\label{crelrat}%
Ratio of NFW $c_{200}$ for relaxed vs.~all $z=0$ halos in various simulations.
}\end{figure}

\begin{figure}
\plotone{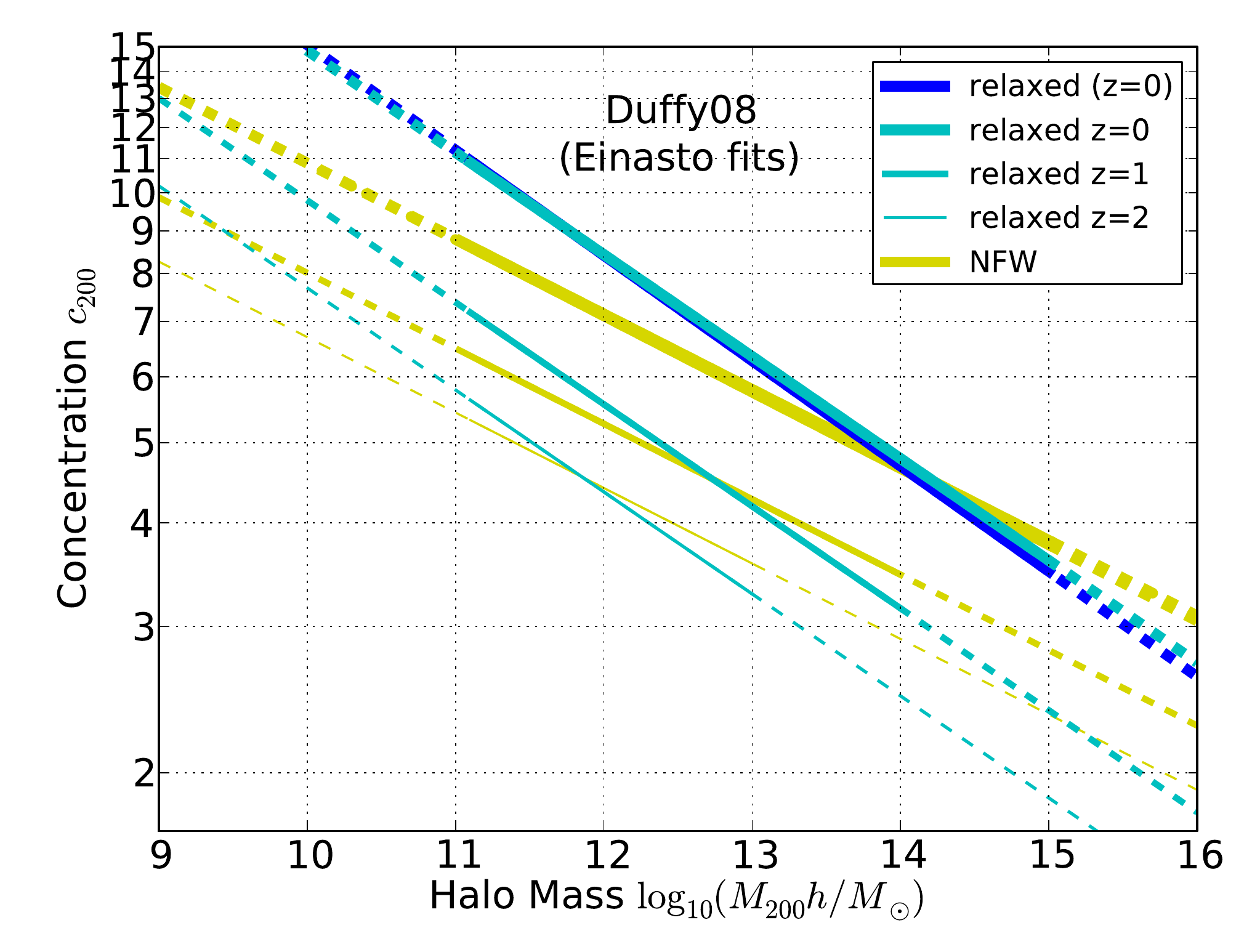}
\caption[]{\label{cMzEinrel}%
Expected concentrations derived from {\it Einasto} profiles compared to NFW profiles.
Plotted are relations for relaxed halos from \cite{Duffy08}.
The yellow NFW lines were plotted in Fig.~\ref{cMzM}.
}\end{figure}

\begin{deluxetable*}{llllrll}  
\tablecaption{\label{tab:cM}
NFW $c_{200}(M_{200})$ fit parameters ($z=0$): 
$c_{200} = c_0 (M_{200} / M_0)^{-\alpha} \pm \Delta c_{200}$}
\tablewidth{0pt}
\tablehead{
\colhead{Sample} &
\colhead{} &
\colhead{Cosmology} &
\colhead{$c_0$} &
\colhead{$M_0 [M_\odot / h]$}  & 
\colhead{$\alpha$}  & 
\colhead{$\Delta \log_{10} c_{200}$}
}
\startdata
Duffy08 & all & WMAP5 &
5.74 & $2\times 10^{12}$ & 0.097 & 0.15\\
Duffy08 & relaxed & WMAP5 &
6.67 & $2\times 10^{12}$ & 0.092 & 0.15\\
Maccio08 & all & WMAP5 & 
6.12 & $10^{12}$ & 0.110 & 0.130\\
Maccio08 & relaxed & WMAP5 & 
6.76 & $10^{12}$ & 0.098 & 0.105\\
Maccio08 & all & WMAP3 & 
5.24 & $10^{12}$ & 0.088 & 0.132\\
Maccio08 & relaxed & WMAP3 & 
5.87 & $10^{12}$ & 0.083 & 0.109\\
Maccio08 & all & WMAP1 & 
7.57 & $10^{12}$ & 0.119 & 0.129\\
Maccio08 & relaxed & WMAP1 & 
8.26 & $10^{12}$ & 0.104 & 0.111\\
Neto07 & all & WMAP1 &
4.67 & $10^{14}$ & 0.11 & 0.094\\
Neto07 & relaxed & WMAP1 & 
5.26 & $10^{14}$ & 0.10 & 0.061\\
\vspace{-0.1in}
\enddata
\end{deluxetable*}

\begin{deluxetable*}{llllrll}  
\tablecaption{\label{tab:cMvir}
NFW $c_{vir}(M_{vir})$ fit parameters ($z=0$): 
$c_{vir} = c_0 (M_{vir} / M_0)^{-\alpha} \pm \Delta c_{vir}$}
\tablewidth{0pt}
\tablehead{
\colhead{Sample} &
\colhead{} &
\colhead{Cosmology} &
\colhead{$c_0$} &
\colhead{$M_0 [M_\odot / h]$}  & 
\colhead{$\alpha$} &
\colhead{$\Delta \log_{10} c_{vir}$}
}
\startdata
Duffy08 & all & WMAP5 &
$7.96\pm0.17$ & $2\times 10^{12}$ & $0.091\pm0.007$ & \nodata\\
Duffy08 & relaxed & WMAP5 &
$9.23\pm0.15$ & $2\times 10^{12}$ & $0.089^{+0.010}_{-0.013}$ & \nodata\\
Maccio08 & all & WMAP5 & 
8.41 & $10^{12}$ & 0.108 & \nodata\\
Maccio08 & relaxed & WMAP5 & 
9.35& $10^{12}$ & 0.094 & \nodata\\
Maccio08 & all & WMAP3 & 
7.26 & $10^{12}$ & 0.086 & \nodata\\
Maccio08 & relaxed & WMAP3 & 
8.22 & $10^{12}$ & 0.080 & \nodata\\
Maccio08 & all & WMAP1 & 
10.26 & $10^{12}$ & 0.114 & \nodata\\
Maccio08 & relaxed & WMAP1 & 
11.25 & $10^{12}$ & 0.099 & \nodata\\
Hennawi07 & all & WMAP1 &  
12.3 & $1.3 \times 10^{13}$ & 0.13 & 0.098\\ 
Gentile07 & all & WMAP3 &  
13.6 & $10^{11}$ & 0.13 & \nodata\\ 
Bullock01 & all & WMAP1 &  
9 & $1.3 \times 10^{13}$ & 0.13 & 0.14\tablenotemark{a}\\ 
Comerford07 & all & observed & 
$14.5\pm 6.4$ & $1.3 \times 10^{13}$ & $0.15\pm 0.13$ & 0.15\\
\vspace{-0.1in}
\enddata
\tablenotetext{a}{\citet[footnote 10]{Wechsler02}
claim that the scatter of $\Delta \log_{10} c_{vir} = 0.18$ reported by Bullock et al. (2001)
was a bit too high and should actually be 0.14, 
thus bringing it in line with their own measured scatter.}
\end{deluxetable*}

\begin{deluxetable*}{lllllrll}  
\tablecaption{\label{tab:cMz}
NFW $c(M,z)$ fit parameters for $0 < z < 2$: 
$c = c_0 (M / M_0)^{-\alpha} (1+z)^{-\beta}$}
\tablewidth{0pt}
\tablehead{
\colhead{Sample} &
\colhead{} &
\colhead{Cosmology} &
\colhead{$\Delta$} &
\colhead{$c_0$} &
\colhead{$M_0 [M_\odot / h]$}  & 
\colhead{$\alpha$} &
\colhead{$\beta$}
}
\startdata
Duffy08 & all & WMAP5 & 200 &
$5.71\pm0.12$ & $2\times 10^{12}$ & $0.084\pm0.006$ & $0.47\pm0.04$\\
Duffy08 & relaxed & WMAP5 & 200 &
$6.71\pm0.12$ & $2\times 10^{12}$ & $0.091\pm0.009$ & $0.44\pm0.05$\\
 Duffy08 & all & WMAP5 & vir &
$7.85^{+0.17}_{-0.18}$ & $2\times 10^{12}$ & $0.081\pm0.006$ & $0.71\pm0.04$\\
Duffy08 & relaxed & WMAP5 & vir &
$9.23^{+0.17}_{-0.16}$ & $2\times 10^{12}$ & $0.090\pm0.009$ & $0.69\pm0.05$\\
\vspace{-0.1in}
\enddata
\end{deluxetable*}



\begin{deluxetable*}{llclrlcrrrl} 
\tablecaption{\label{tab:cMobs}
Observed $c(M)$ fit parameters ($z=0$): $c = c_0 (M / M_0)^{-\alpha} (1+z)^{-\beta}$}
\tablewidth{0pt}
\tablehead{
\colhead{} &
\colhead{} &
\colhead{} &
\colhead{} &
\colhead{$M_0$} &
\colhead{} &
\colhead{} &
\colhead{} &
\colhead{$M_{\rm min}$} &
\colhead{$M_{\rm max}$} &
\colhead{}\\
\colhead{Sample} &
\colhead{Analysis} &
\colhead{$\Delta_c$} &
\colhead{$c_0$} &
\colhead{[$h^{-1}  M_\odot$]} &
\colhead{$\alpha$} &
\colhead{$\Delta \log_{10}c$} &
\colhead{N} &
\colhead{[$h^{-1}  M_\odot$]} &
\colhead{[$h^{-1}  M_\odot$]} &
\colhead{$z$}
}
\startdata
Comerford07\tablenotemark{b} & compilation & vir &
$14.5\pm 6.4$ & $1.3 \times 10^{13}$ & $0.15\pm 0.13$ & 0.15 &
62 & $5 \times 10^{13}$ & $4 \times 10^{15}$ & $0.003-0.89$ \\
Buote07 & X-ray & vir &
$9.0\pm 0.4$ & $10^{14}$ & $0.172\pm 0.026$ & \nodata &
39 relaxed & $6 \times 10^{12}$ & $2 \times 10^{15}$ & $0.0033-0.2302$\\
SchmidtAllen07 & X-ray & vir &
$7.55\pm 0.90$\tablenotemark{c} & $8 \times 10^{14}$ & $0.45\pm 0.12$\tablenotemark{c} & \nodata &
34 relaxed & $2 \times 10^{14}$ & $4 \times 10^{15}$ & $0.06-0.$7\\
Okabe09 & WL & vir &
$8.45^{+3.91}_{-2.80}$ & $10^{14}$ & $0.41\pm 0.19$ & 0.19 &
30 & $2 \times 10^{14}$ & $1.5 \times 10^{15}$ & $0.15-0.30$\\
Johnston07 & stacked WL & 200 &
$4.1\pm 1.2$ & $1.3 \times 10^{13}$ & $0.12\pm 0.04$ &  \nodata &
130,000 & $5 \times 10^{12}$ & $5 \times 10^{14}$ & $\sim0.25$\\
Mandelbaum08\tablenotemark{d} & stacked WL & 54 &
$4.6\pm 0.7$ & $10^{14}$ & $0.13\pm 0.07$ & \nodata &
222,699 & $3 \times 10^{13}$ & $6 \times 10^{14}$ & $\sim0.22$\\
Mandelbaum08\tablenotemark{e} & stacked WL & 200 &
$2.5\pm 0.4$ & $6 \times 10^{13}$ & $0.13\pm 0.07$ & \nodata &
222,699 & $3 \times 10^{13}$ & $6 \times 10^{14}$ & $\sim0.22$\\
\vspace{-0.1in}
\enddata
\tablecomments{SchmidtAllen07 find $\beta = 0.71\pm 0.52$\tablenotemark{c}, 
but all others fix $\beta = 1$ while fitting only $(c_0, \alpha)$.}
\tablenotetext{b}{Includes Buote07 and SchmidtAllen07}
\tablenotetext{c}{Quoted uncertainties are 95\% rather than 1-$\sigma$}
\tablenotetext{d}{Mandelbaum08 used $\Delta_{vir} = 200$ 
({\it not} $\Delta_c = 200$)
and assumed $\Omega_m = 0.27$, corresponding to $\Delta_c = \Omega_m \Delta_{vir} = 54$.}
\tablenotetext{e}{Converted from previous line}
\end{deluxetable*}



\begin{deluxetable}{lcccrrr}  
\tablecaption{\label{tab:sims}
Simulations
}
\tablewidth{0pt}
\tablehead{
\colhead{Simulation} &
\colhead{Cosmology} &
\colhead{$M_{\rm min}$} &
\colhead{$M_{\rm max}$} &
\colhead{N} &
\colhead{N} &
\colhead{particles}\\
\colhead{} &
\colhead{(see Table \ref{tab:cosmo})} &
\colhead{[$h^{-1}  M_\odot$]} &
\colhead{[$h^{-1}  M_\odot$]} &
\colhead{all} &
\colhead{relaxed} &
\colhead{within $r_{vir}$}
}
\startdata
Duffy08   &
WMAP5 & $10^{11}$ & $10^{15}$ & 1,269 & 561 & 10,000\\
Maccio08 & 
WMAP5 & $10^{10}$ & $10^{15}$ & 9,988 & 7,060 & 500\\
Neto07 (Millennium) & 
WMAP1 & $10^{12}$ & $10^{15}$ & 53,626 & 39,330 & 10,000\\
Hennawi07 & 
WMAP1 & $10^{14}$ & $10^{15}$ & 878 & \nodata & \nodata\\
Bullock01 & 
WMAP1 & $10^{11}$ & $10^{14}$ & $\sim 5,000$ & \nodata & 150--120,000\\ 
Gentile07 (NFW96) & 
WMAP3 & $3 \times 10^{11}$ & $3 \times 10^{15}$ & 19 & \nodata & 5,000--10,000\\
\vspace{-0.1in}
\enddata
\end{deluxetable}

\begin{deluxetable}{llll}  
\tablecaption{\label{tab:cosmo}
Cosmological parameters}
\tablewidth{0pt}
\tablehead{
\colhead{Author} &
\colhead{WMAP} &
\colhead{$\Omega_m$} &
\colhead{$\sigma_8$}
}
\startdata
Bullock01 & WMAP1 & 0.3 & 1.0\\
Hennawi07 & WMAP1 & 0.3 & 0.95\\
Neto07: Millennium & WMAP1 & 0.25 & 0.90\\
Maccio08 & WMAP1 & 0.268 & 0.90\\ 
Maccio08 & WMAP3 & 0.238 & 0.75\\
Maccio08, Duffy08 & WMAP5 & 0.258 & 0.796\\
\nodata & WMAP7 & 0.26 & 0.803\\
\vspace{-0.1in}
\enddata
\end{deluxetable}





\addtocounter{figure}{-9}


\addtocounter{figure}{8}

\begin{figure}
\plotone{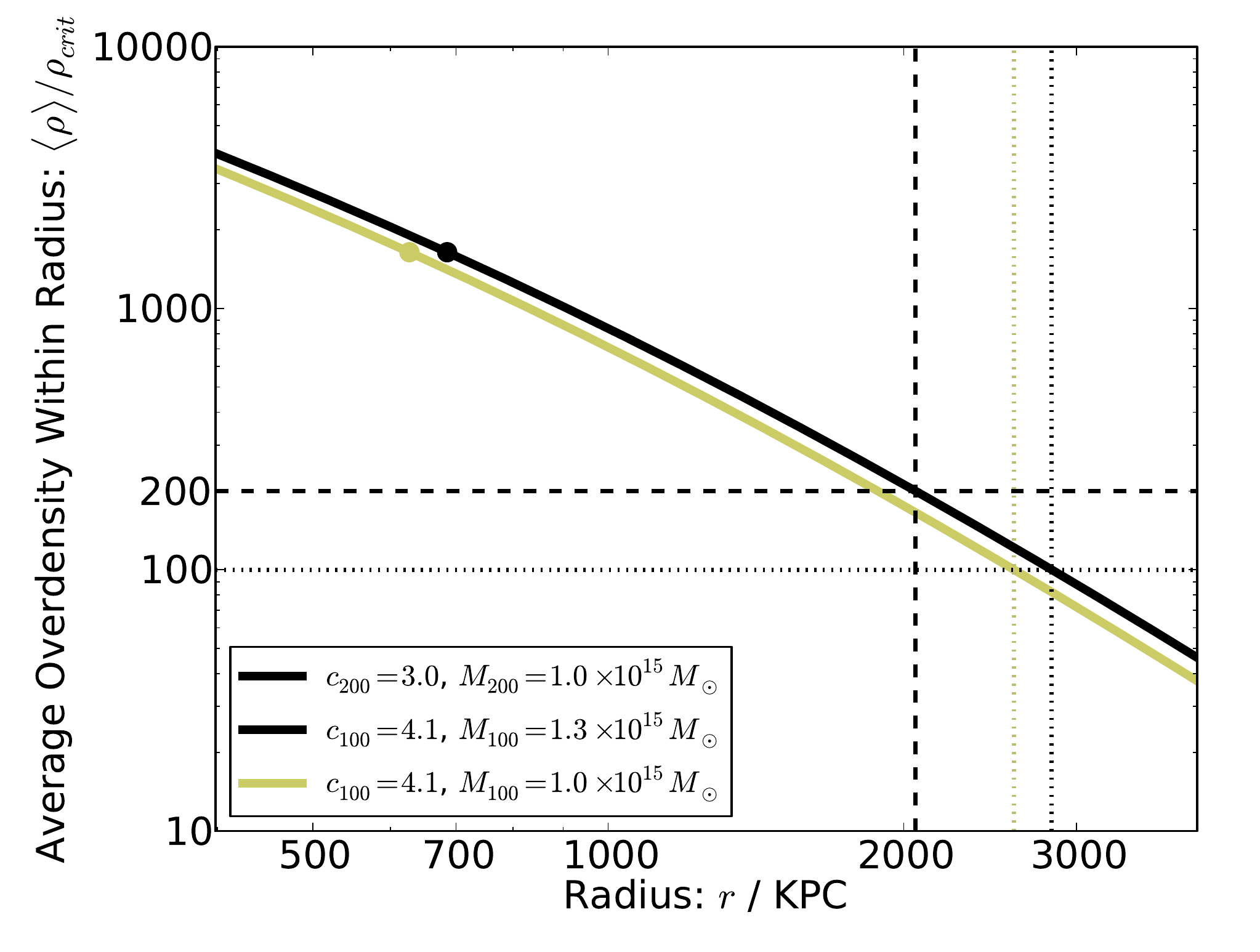}
\caption[]{\label{fig:ccompareoverdens}%
For a fixed NFW curve $(r_s, \rho_s)$,
illustration of conversion of $(c,M)$ between different values of $\Delta_c$.
From Fig.~\ref{virrat}, we find 
$c_{100} \approx 1.37 c_{200}$ and 
$M_{100} \approx 1.3 M_{200}$.
The two (perfectly overlapping) black curves have identical $(r_s, \rho_s)$.
Dots mark $r_s$, while the dashed lines mark the overdensities and $r_{vir}$
for $\Delta_c = 200$ and 100.
Note that for fixed $c$, $r_s$ and $r_{vir}$ vary with $M_{vir}$ (black vs.~yellow).
}\end{figure}

\begin{figure}
\plotone{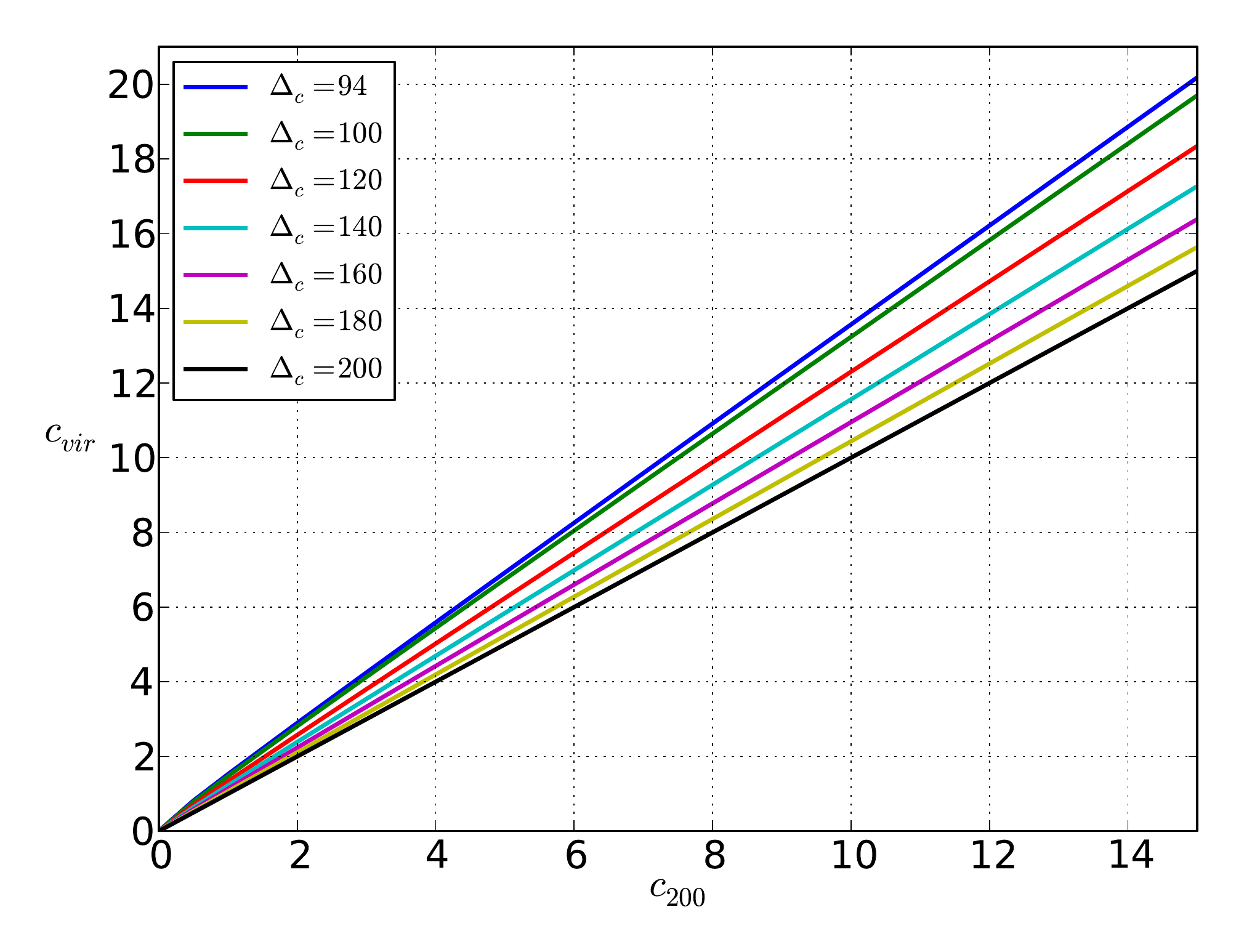}
\caption[]{\label{fig:cconv}%
Conversion from NFW $c_{200}$ to $c_{vir}$ 
for $\Delta_c = 94,100,120,140,160,180,200$.
The relations are extremely linear for $c_{200} > 2$.
}\end{figure}


\begin{figure*}
\plottwo{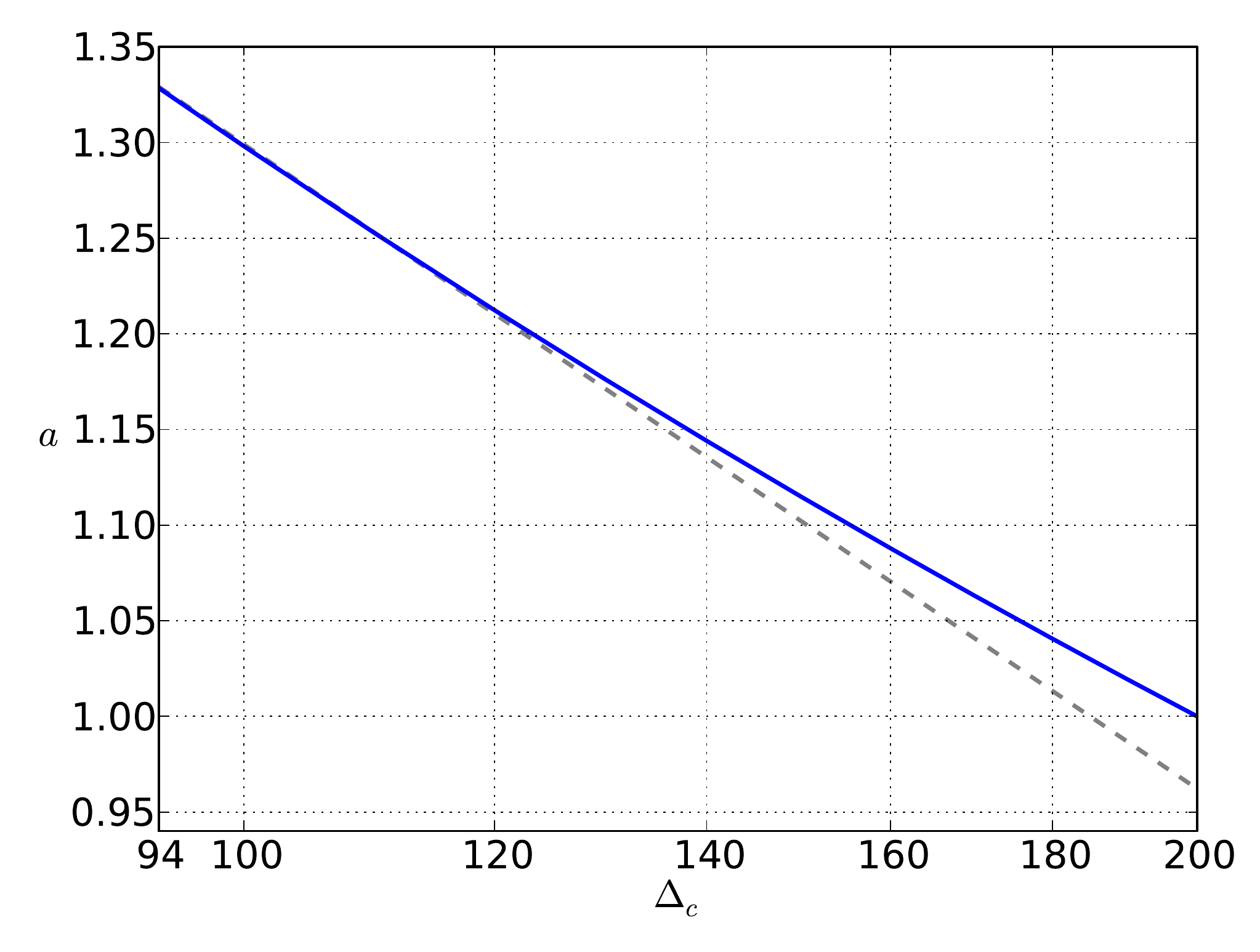}{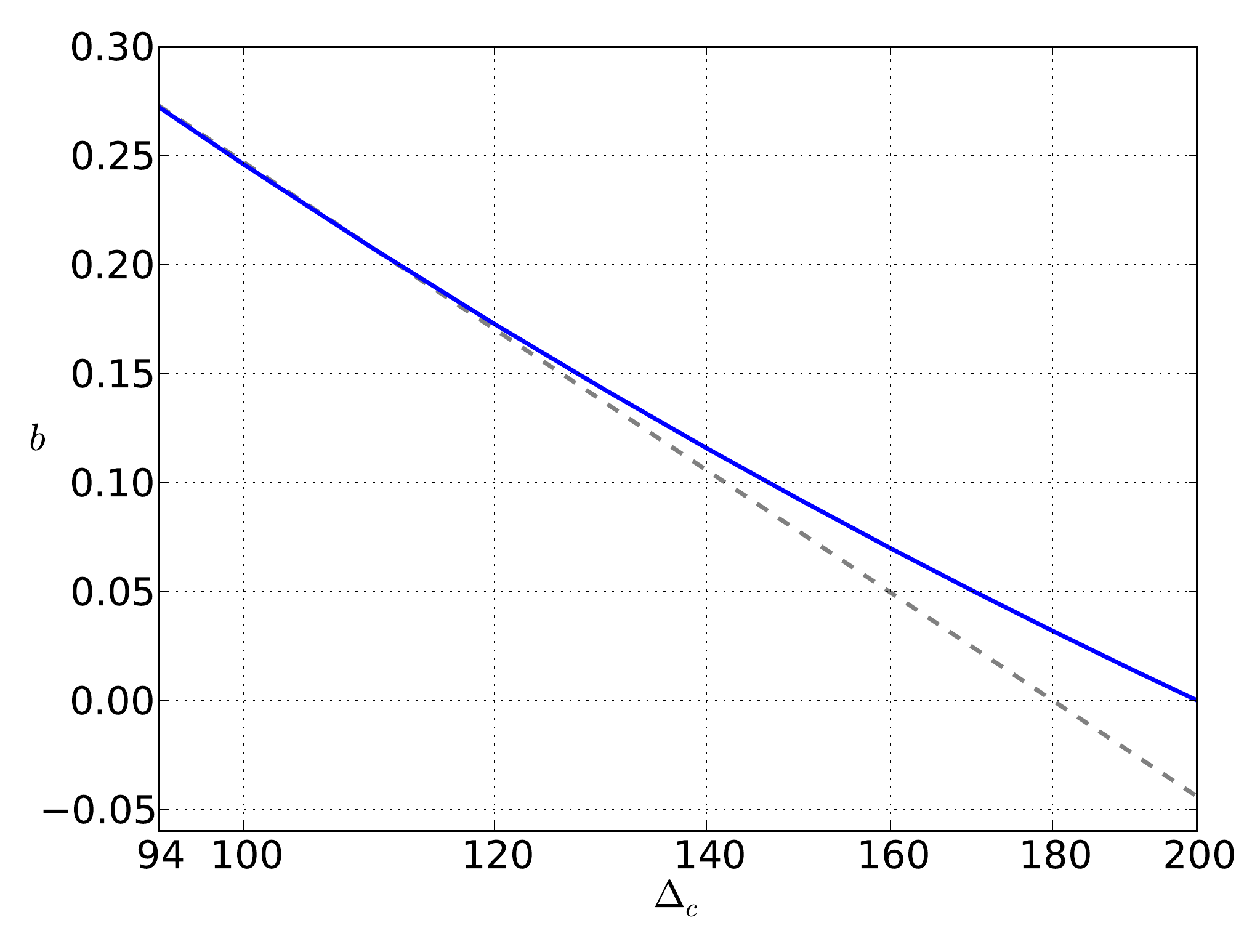}
\caption[]{\label{fig:Dc}%
Parameters $a$ and $b$ for conversion of NFW $c_{vir} \approx a \thinspace c_{200} + b$.
Dashed lines are the relations given in Eqs.~\ref{cconva} \& \ref{cconvb}.
}\end{figure*}

\begin{figure*}
\plottwo{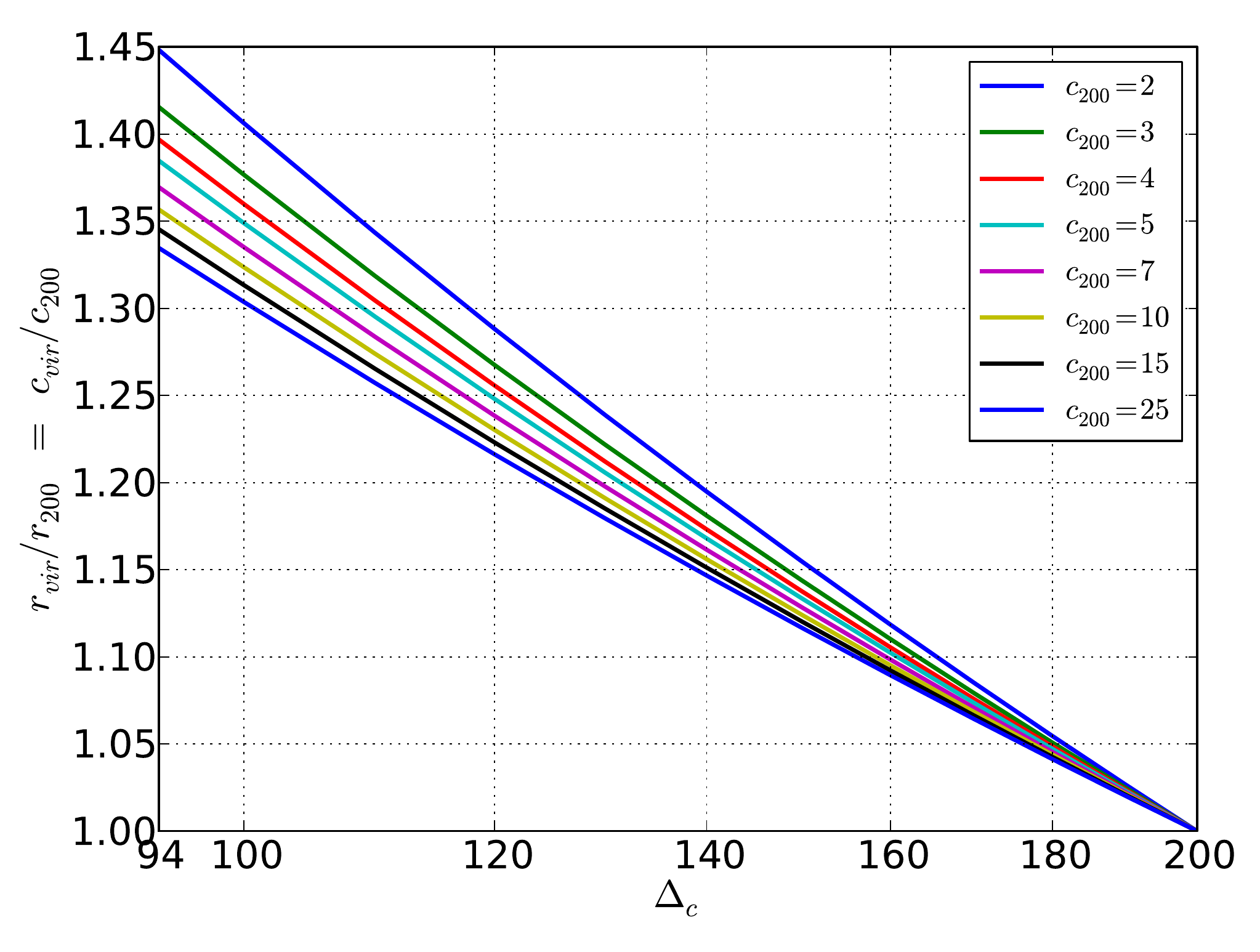}{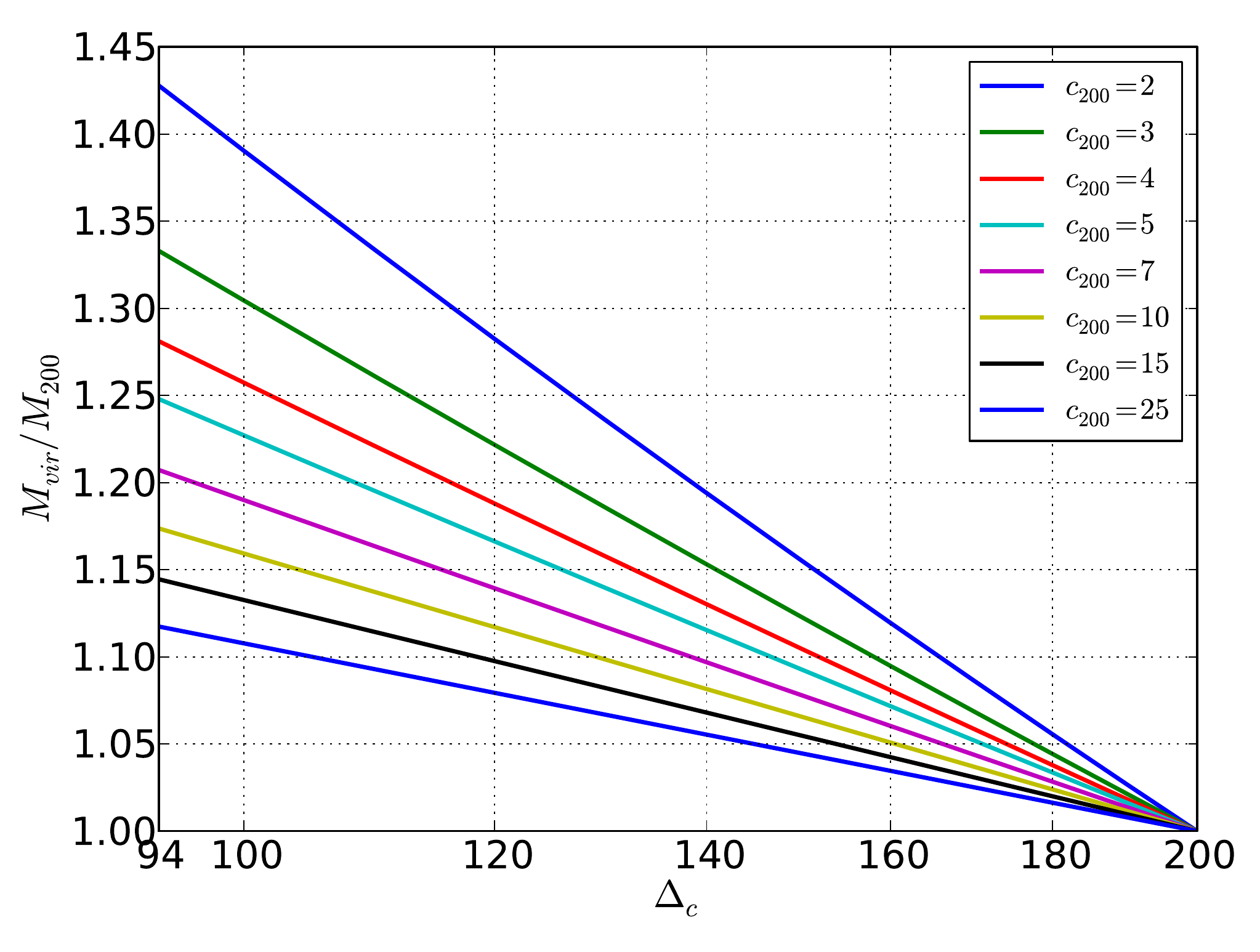}
\caption[]{\label{virrat}%
Ratios of $r_{vir}$ (left) and $M_{vir}$ (right)
to the $\Delta_c = 200$ values
as a function of NFW ($c$, $\Delta_c$).
Note that $r_{vir} / r_{200} = c_{vir} / c_{200}$
since we hold $r_s$ (and $\rho_s$) fixed
and we define $r_{vir} = c_{vir} r_s$ and $r_{200} = c_{200} r_s$.
}\end{figure*}


\begin{figure*}
\plottwo{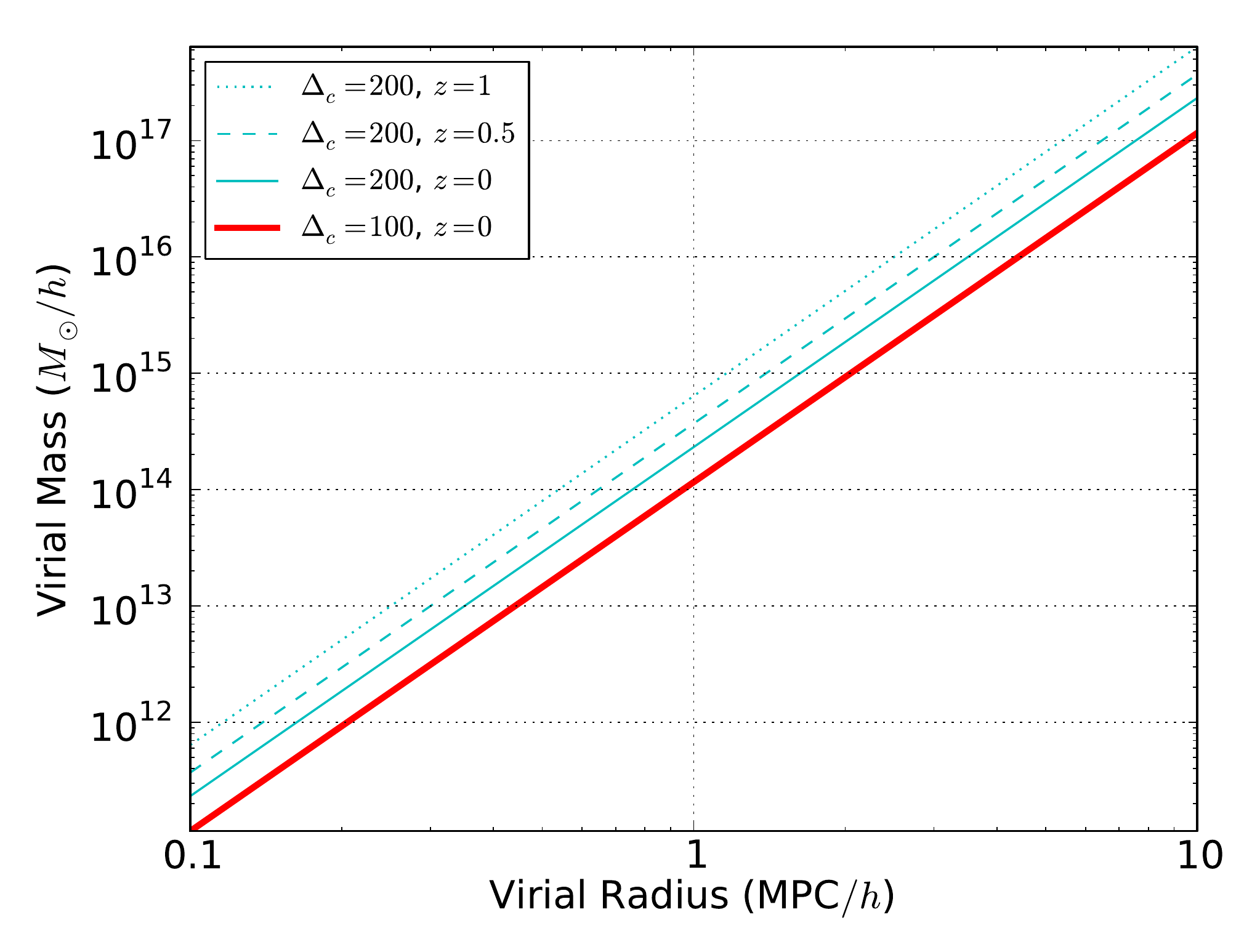}{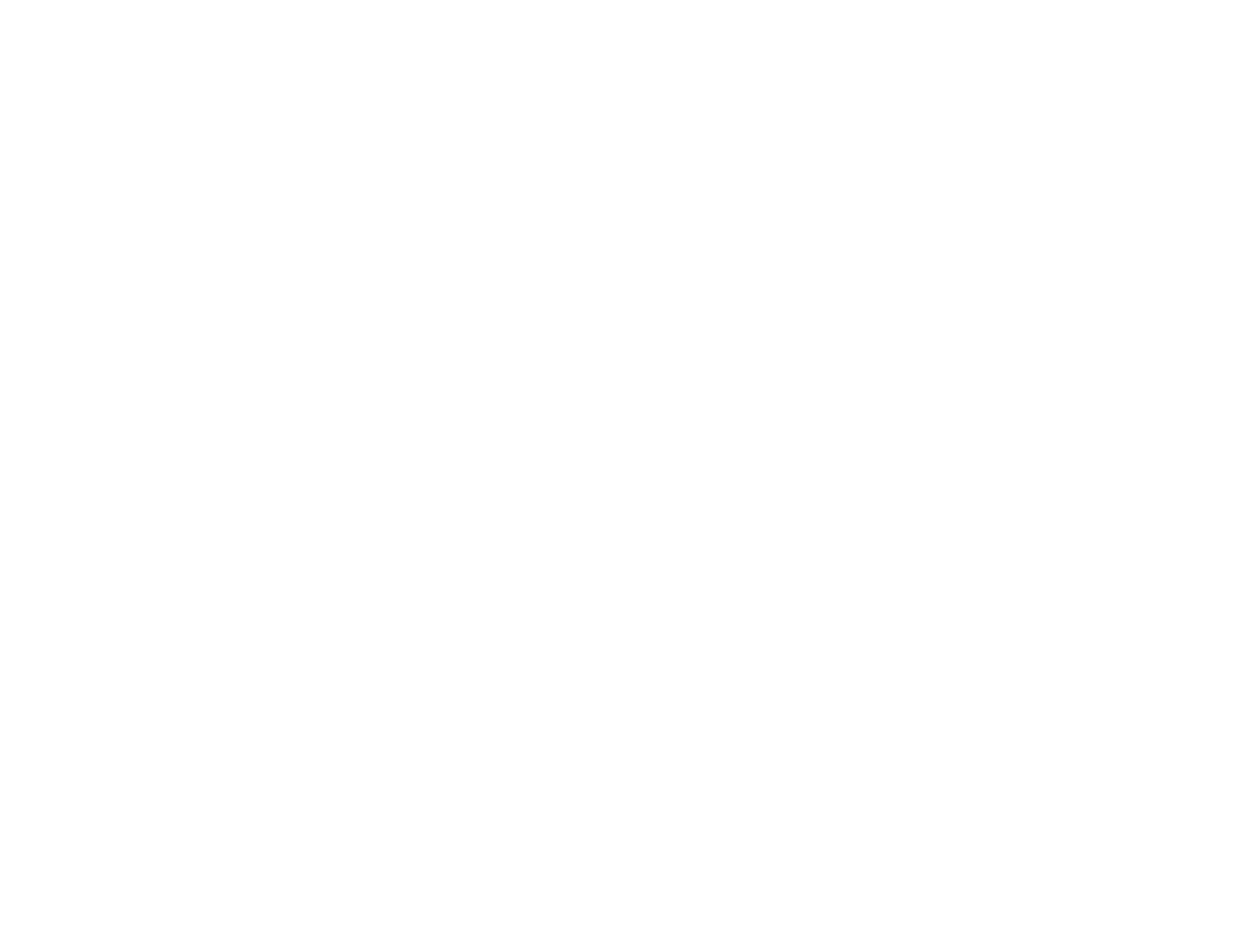}
\caption[]{\label{Mrvir}%
The relation between $r_{vir}$ and $M_{vir}$ is independent of the assumed profile,
only dependent on the chosen overdensity, cosmology, and halo redshift (Eq. \ref{eq:Mvir}).
For a $z = 0$ halo with $r_{vir} = 1~{\rm Mpc} / h$, $M_{vir} = 1.16 \times 10^{14} M_\odot / h ~ (\Delta_c / 100)$.
}\end{figure*}


\begin{figure*}
\plottwo{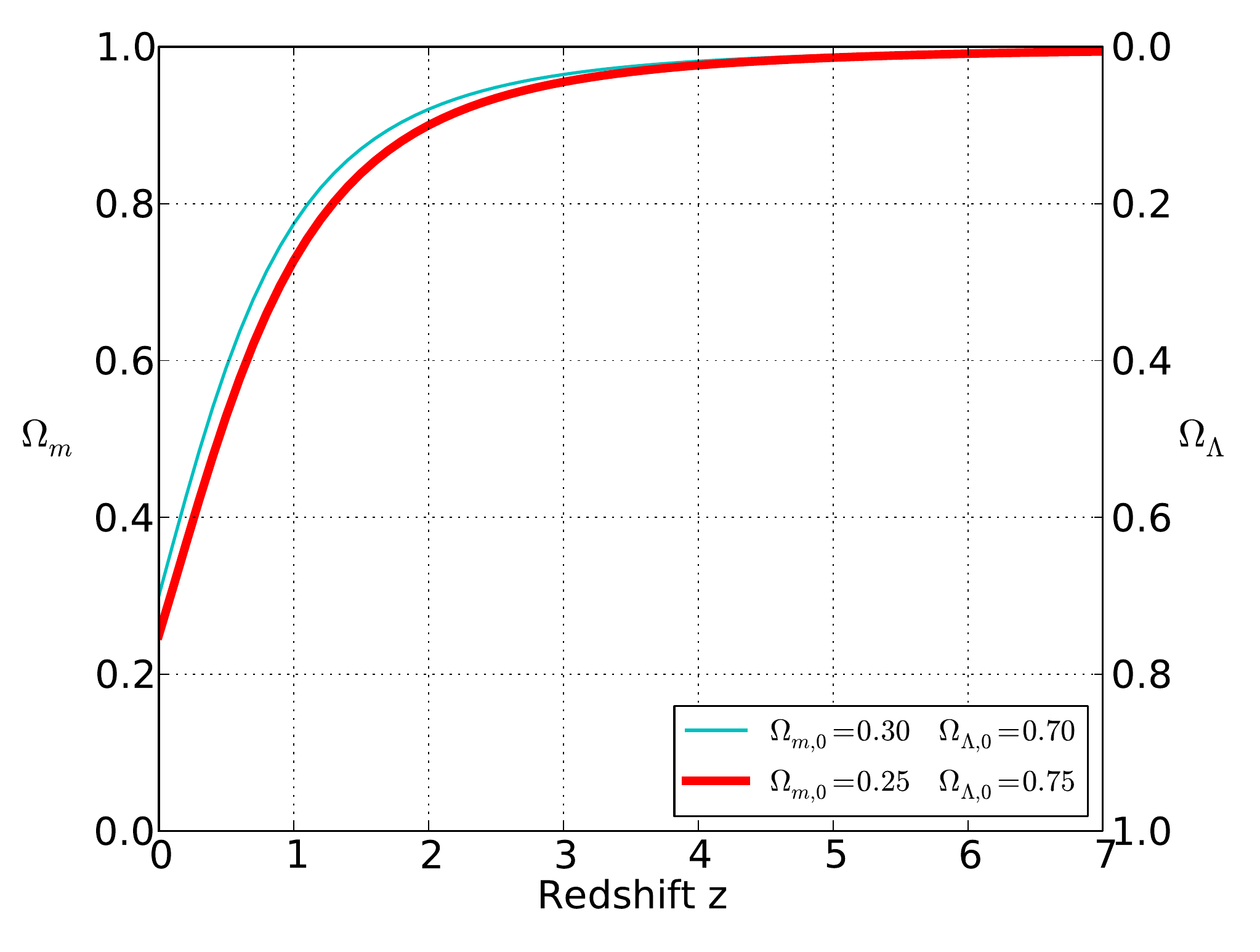}{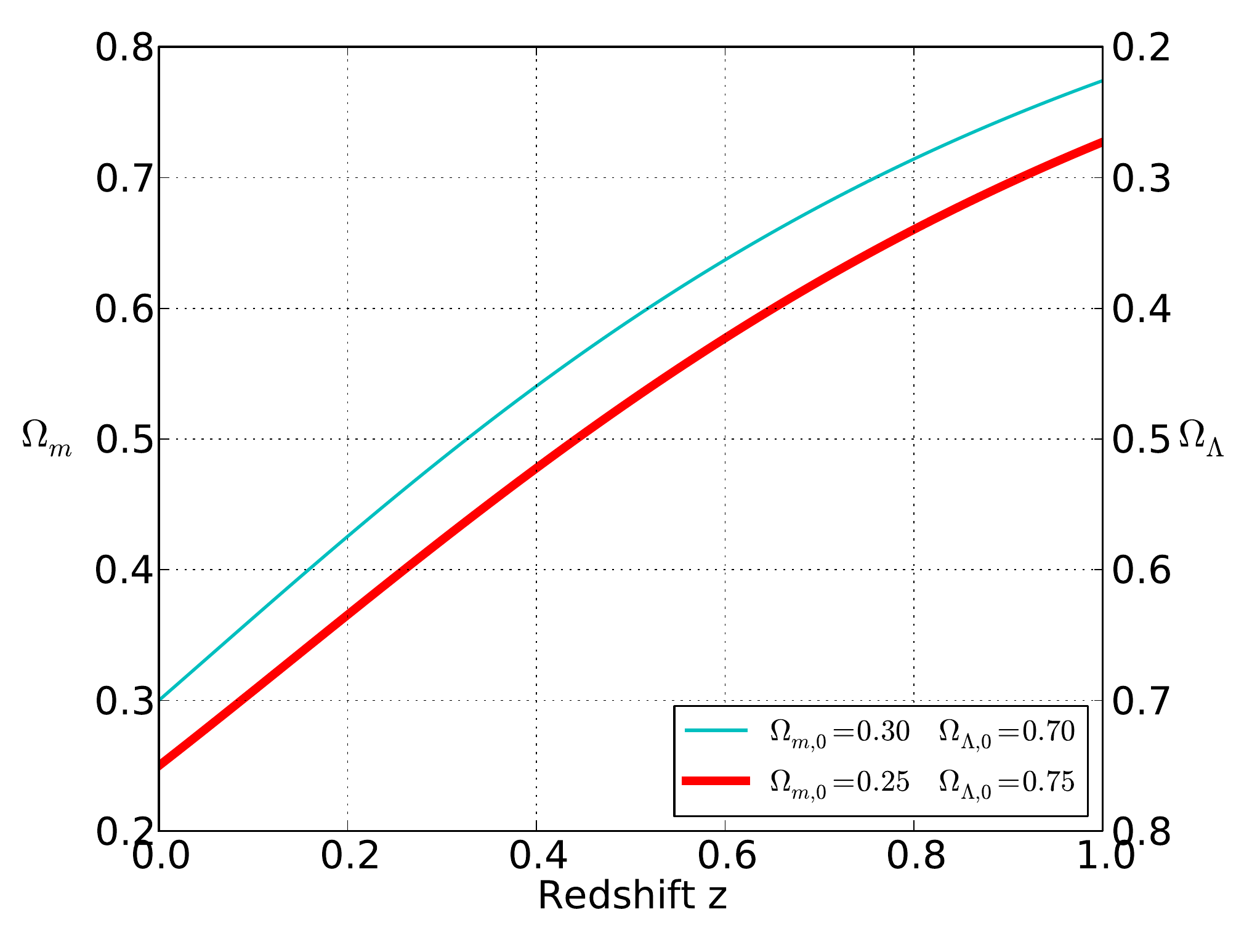}
\caption[]{\label{OmOLz}%
$\Omega_m(z) = 1 - \Omega_\Lambda(z)$ in a flat universe with a cosmological constant
}\end{figure*}

\begin{figure*}
\plottwo{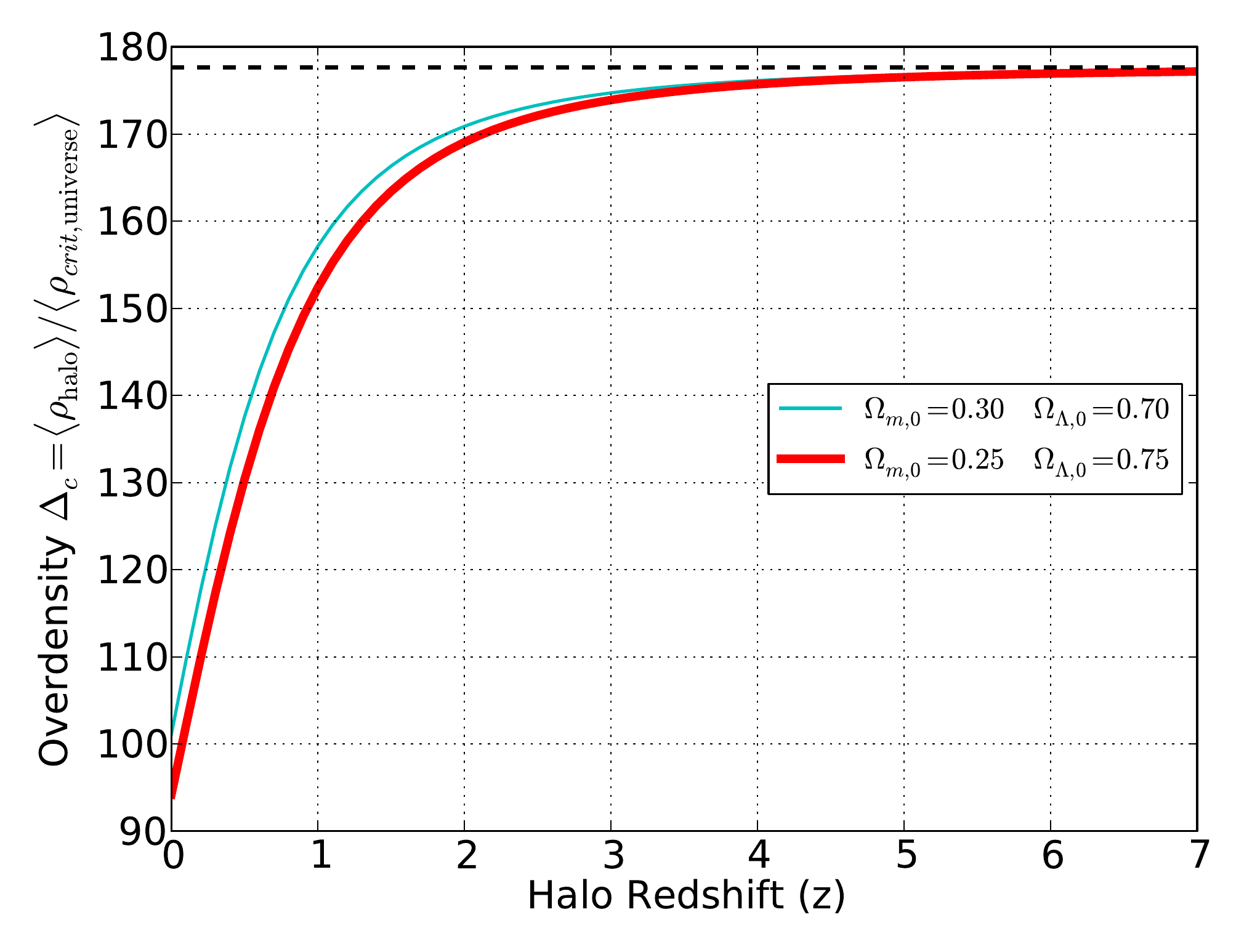}{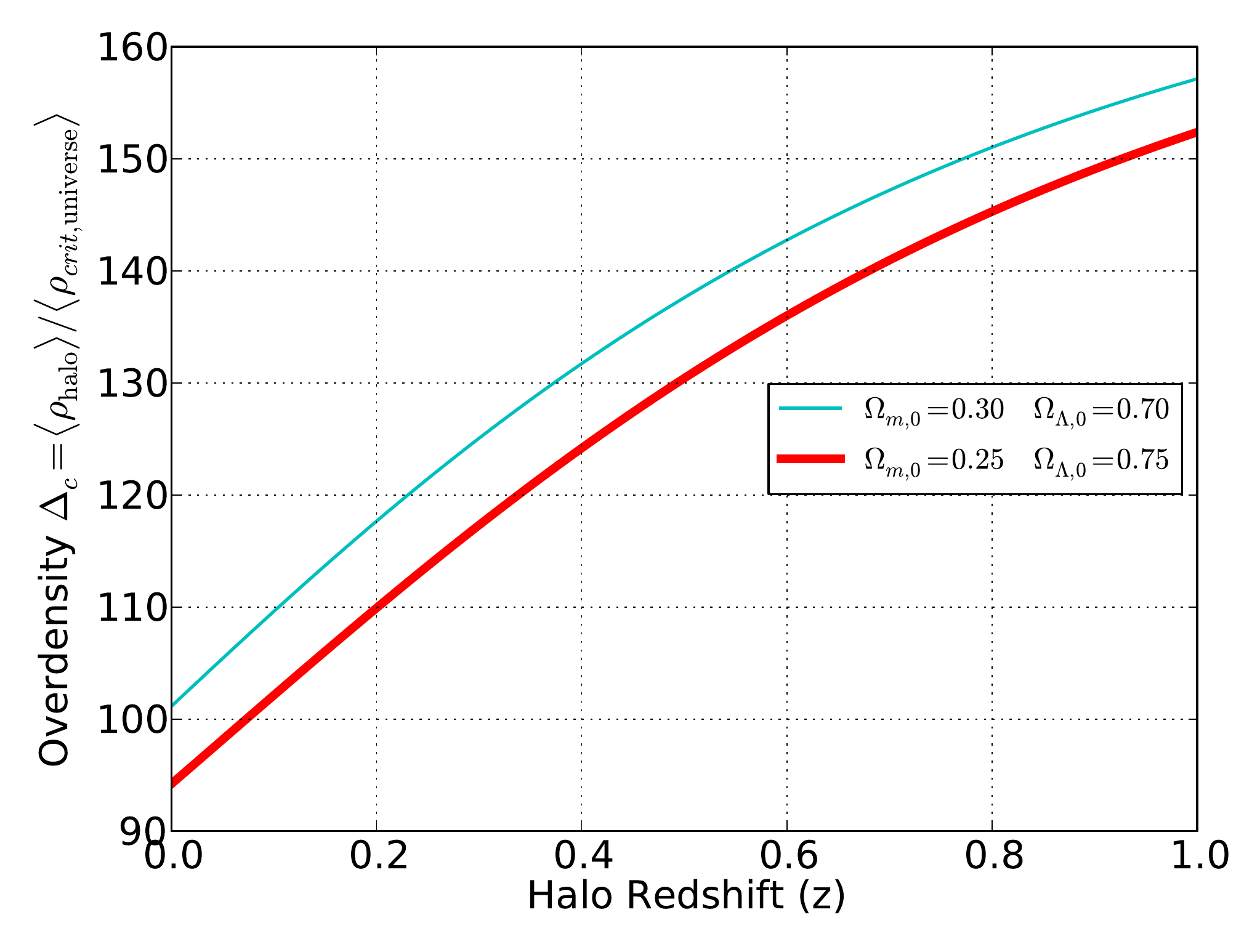}
\caption[]{\label{dc}%
$\Delta_c(z)$: The average overdensity above $\rho_{crit}$ within a collapsed halo
}\end{figure*}

\begin{figure*}
\plottwo{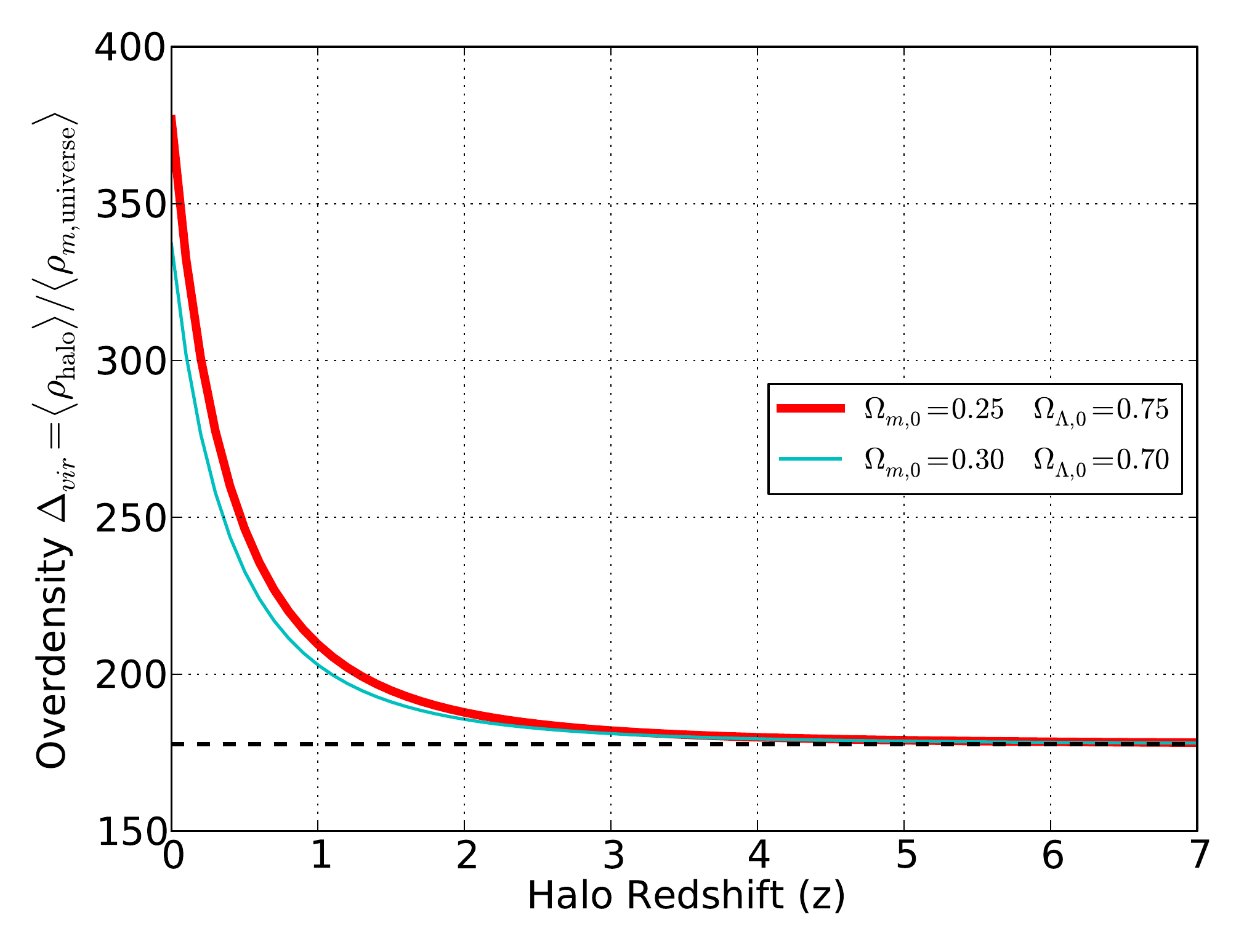}{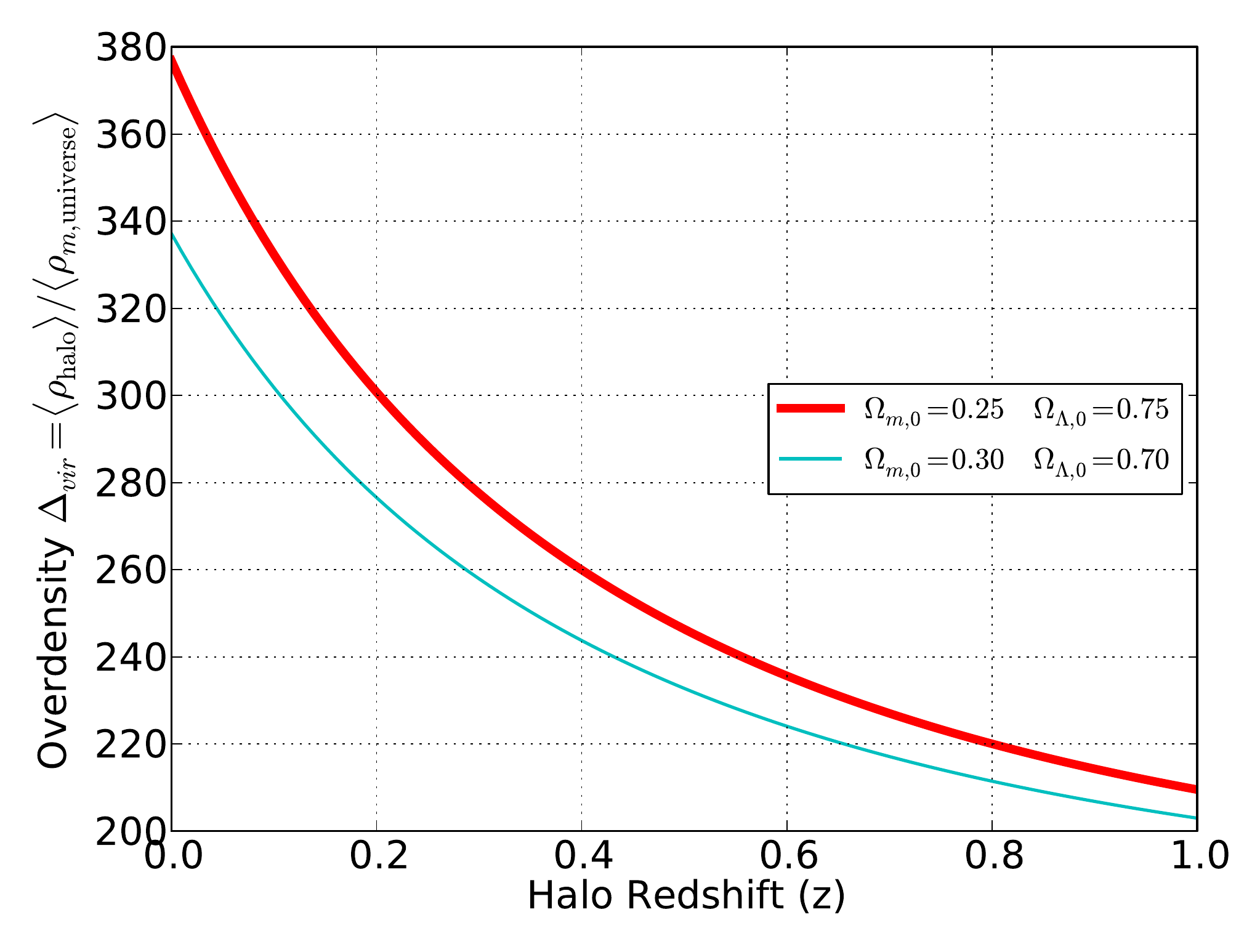}
\caption[]{\label{dcvir}%
$\Delta_{vir}(z)$: The average overdensity above $\rho_m = \Omega_m \rho_{crit}$ within a collapsed halo
}\end{figure*}


\begin{figure}
\plotone{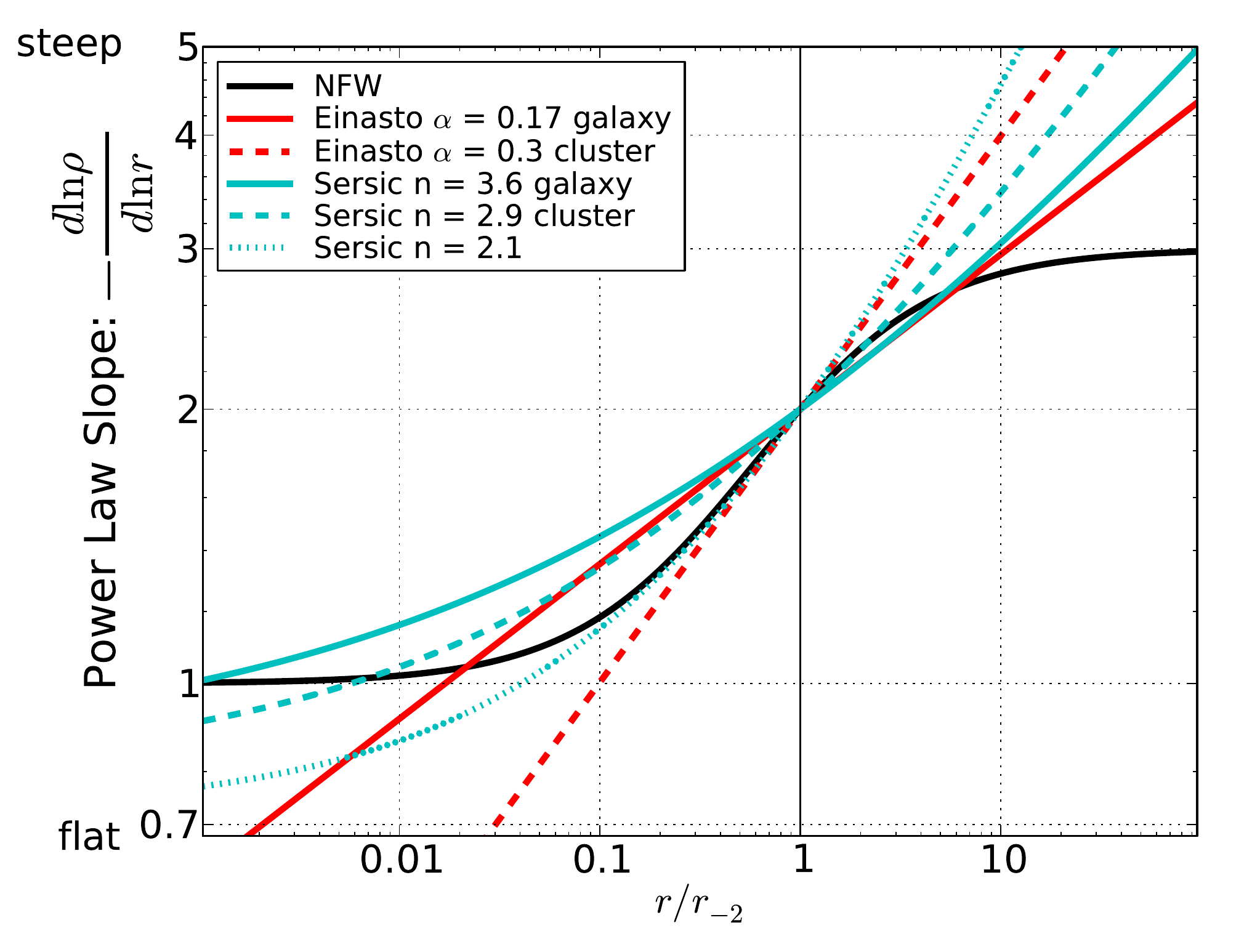}
\caption[]{\label{powerlaws}%
Power law slopes $\gp$ for 3-D density $\rho \propto r^{-\gp}$.
Here we compare the NFW, Einasto, and S\'ersic profiles.
For S\'ersic we use the Prugniel-Simien approximation.
A realistic range of $\alpha$ and $n$ are plotted for the profiles,
as found in simulations \citep{Merritt06,Gao08,Navarro10}.
}\end{figure}

\begin{figure}
\plotone{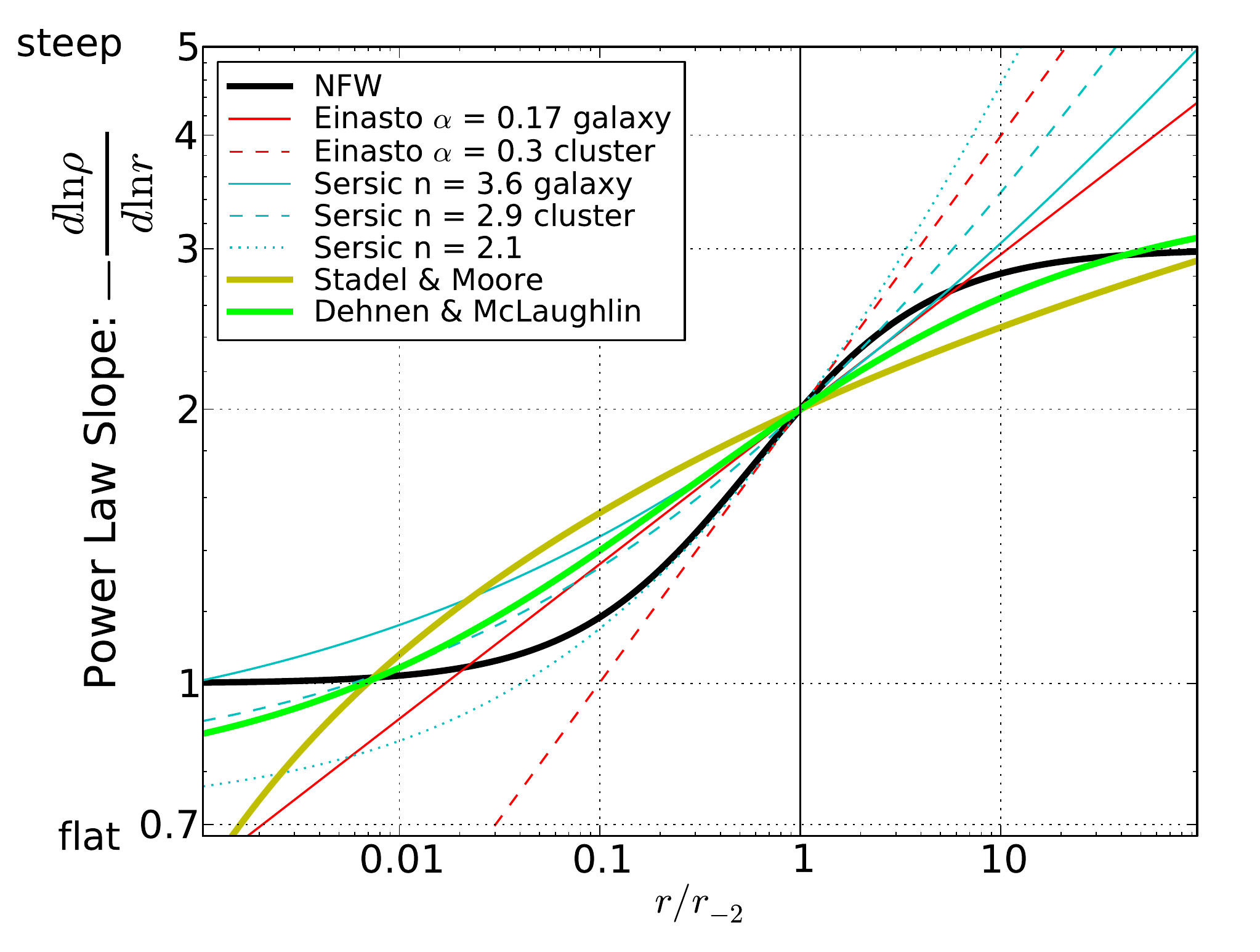}
\caption[]{\label{powerlaws_SMDM}%
Power law slopes $\gp$ for 3-D density $\rho \propto r^{-\gp}$.
In this figure we add the Dehnen-McLaughlin and Stadel-Moore profiles
with $\gamma = 7/9$ and $\lambda = 0.10$, respectively.
}\end{figure}

\begin{figure}
\plotone{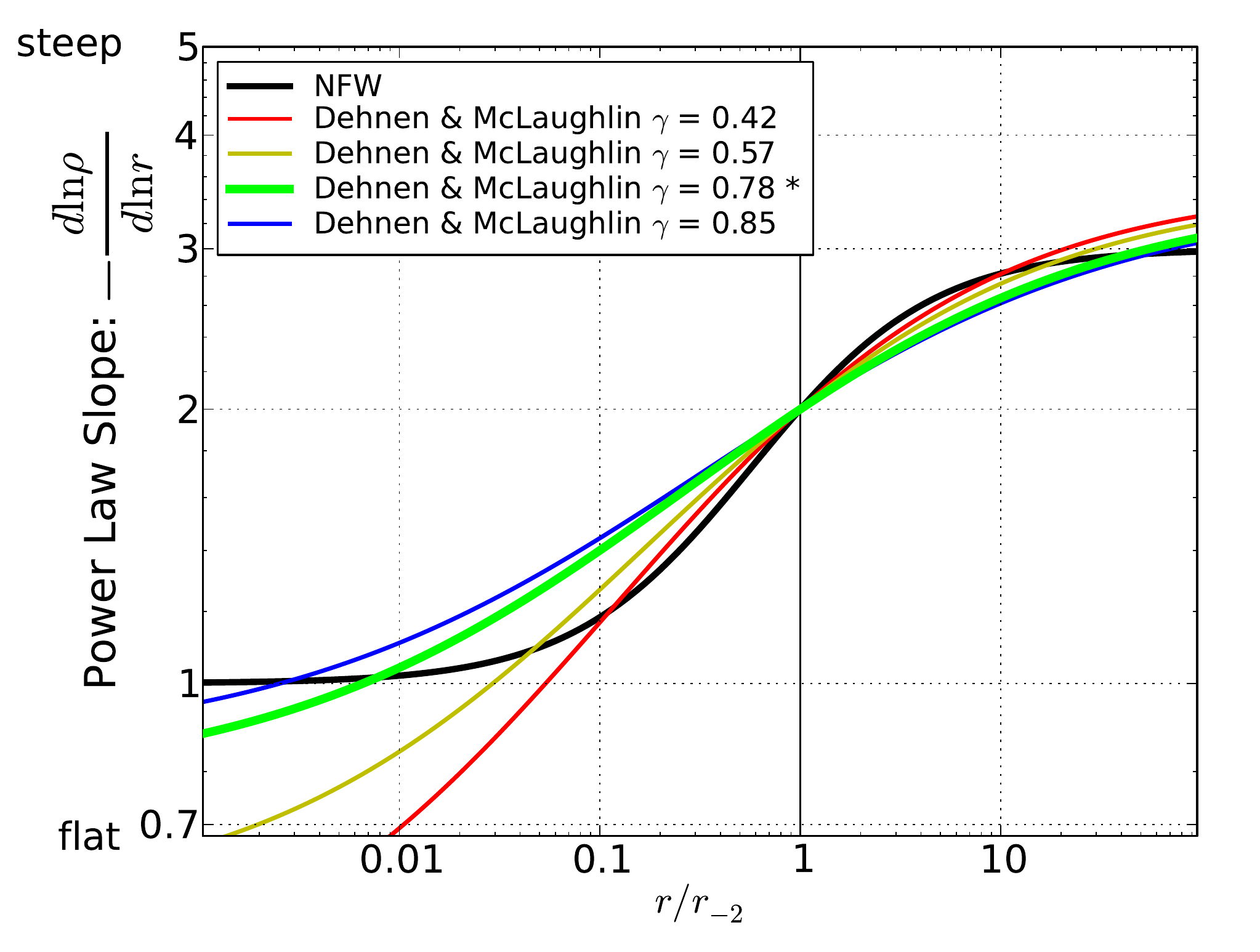}
\caption[]{\label{powerlaws_DM}%
Power law slopes $\gp$ for 3-D density $\rho \propto r^{-\gp}$.
Here we compare various values of $\gamma$ for Dehnen-McLaughlin
as observed in simulations \citep{Merritt06}.
The value marked with a * (7/9) is used in their 2-parameter fitting formula.
}\end{figure}

\begin{figure}
\plotone{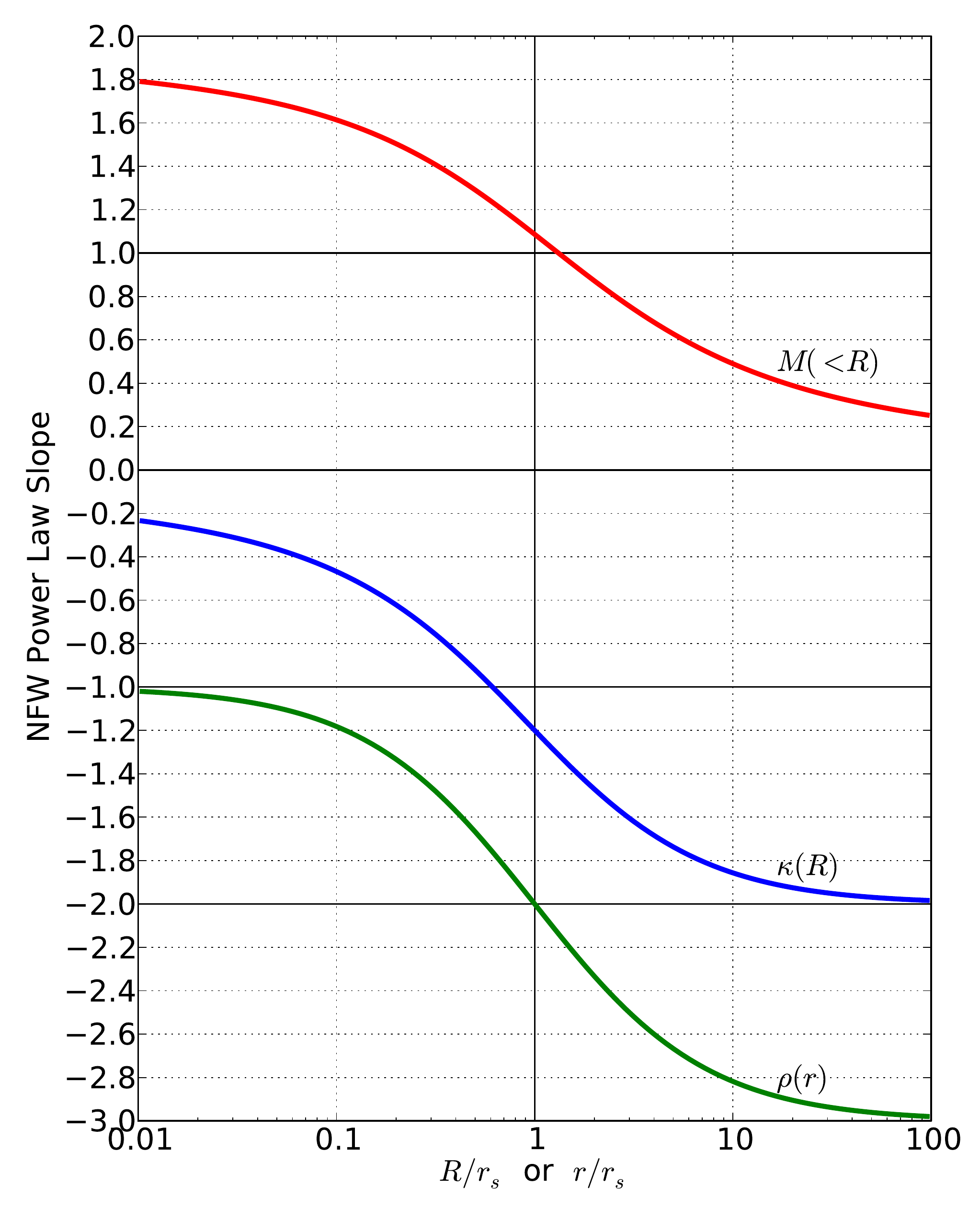}
\caption[NFW fits]{\label{NFWpowerlaws}%
NFW power law slopes for
projected mass within a cylinder $M(<R)$,
projected mass density $\kappa(R)$,
and 3-D density $\rho(r)$.
}\end{figure}

\clearpage
\bibliographystyle{astroads}
\bibliography{paperstrunc}



\end{document}